\documentclass[preprint,aps,prd,amsmath,amssymb]{revtex4}
\usepackage{graphicx}
\usepackage{color}
\usepackage{graphicx}
\usepackage{dcolumn}
\usepackage{bm}
\usepackage{amsmath}
\usepackage{amsfonts}
\usepackage{bbm}
\usepackage{subfigure}
\usepackage{setspace}
\usepackage{slashed}
\newcommand{\beq}{\begin{eqnarray}}
\newcommand{\eeq}{\end{eqnarray}}

\newcommand{\bmp}{\noindent\begin{minipage}{16cm}}
\newcommand{\emp}{\end{minipage}\vskip 7mm} 

\def\drawbox#1#2{\hrule height#2pt
        \hbox{\vrule width#2pt height#1pt \kern#1pt
              \vrule width#2pt}
              \hrule height#2pt}
\def\Fund#1#2{\vcenter{\vbox{\drawbox{#1}{#2}}}}
\def\Asym#1#2{\vcenter{\vbox{\drawbox{#1}{#2}
              \kern-#2pt 
              \drawbox{#1}{#2}}}}

\def\fund{\Fund{6.4}{0.3}}

\begin{document}
\title{\Large  Minimal Walking Technicolor:  \\Set Up for Collider Physics}
\author{Roshan {\sc Foadi}}
\email{roshan@fysik.sdu.dk}
\author{Mads T. {\sc Frandsen}}
\email{toudal@nbi.dk}
\author{Thomas A. {\sc Ryttov}}
\email{ryttov@nbi.dk}
\author{Francesco {\sc Sannino}}
\email{sannino@fysik.sdu.dk}
\affiliation{CERN Theory Division, CH-1211 Geneva 23, Switzerland}
\affiliation{University of Southern Denmark, Campusvej 55, DK-5230 Odense M, Denmark. \\
Niels Bohr Institute, Blegdamsvej 17, DK-2100
Copenhagen, Denmark.}


\begin{abstract}
Different theoretical and phenomenological aspects of the Minimal and Nonminimal Walking Technicolor theories have recently been studied. The goal here is to make the models ready for collider phenomenology. We do this by constructing the low energy effective theory containing scalars, pseudoscalars, vector mesons and other fields predicted by the minimal walking theory. We construct their self-interactions and interactions with standard model fields. Using the Weinberg sum rules, opportunely modified to take into account the walking behavior of the underlying gauge theory, we find interesting relations for the spin-one spectrum. We derive the electroweak parameters using the newly constructed effective theory and compare the results with the underlying gauge theory.  Our analysis is sufficiently general such that the resulting model can be used to represent a generic walking technicolor theory not at odds with precision data.
\end{abstract}

\maketitle

\section{Introduction}
Recently we have uncovered the phase diagram of strongly coupled theories \cite{Dietrich:2006cm,Sannino:2004qp} in the ladder approximation \cite{Appelquist:1988yc,Cohen:1988sq}, as a function
of the number of flavors and colors, with Dirac fermions transforming according to a given
but arbitrary representation of the underlying SU(N) gauge group. Further studies of the conformal window and its properties can be found in \cite{Appelquist:1996dq,MY,Gies:2005as}.

 We have also identified a number of strongly coupled theories which can dynamically break the electroweak gauge symmetry and pass the precision tests. The systematic analysis started in \cite{Dietrich:2005jn} and was extended in \cite{Dietrich:2006cm}. A related study can be found in \cite{Christensen:2005cb}. First principle numerical lattice computations have already initiated the study of the proposed conformal window, and preliminary results just appeared \cite{Catterall:2007yx} which seem to support the near conformal (walking) behavior of the simplest theory. These are initial investigations on very small lattices which encourage one to embark on
a more serious study on larger lattices. The simplest of these theories has fermions in the two index symmetric representation of the gauge group, and is argued to already walk with only two Dirac flavors and two colors \cite{Sannino:2004qp}. Therefore, when used to break the electroweak symmetry dynamically, we will call it the {\it Minimal Walking Technicolor} (MWT) model. This extension of the standard model passes the electroweak precision constraints \cite{Dietrich:2005jn} while displaying some interesting  features. {}For example, it allows for  
a successful unification of the standard model couplings at the one loop level \cite{Gudnason:2006mk}. Also interesting
 types of dark matter components \cite{Kainulainen:2006wq,Kouvaris:2007iq,Gudnason:2006yj} can  be envisioned. 
 The walking dynamics was first introduced in \cite{Eichten:1979ah,Holdom:1981rm,Yamawaki:1985zg,Appelquist:an,Lane:1989ej} to explain the breaking of the electroweak theory. 
  
Here we take our phenomenological program one step closer to collider phenomenology by constructing the theory at the underlying and effective Lagrangian level. We include the relevant fields which can be discovered at collider experiments, with their self interactions as well as the interactions with the standard model fields. We provide the link with the underlying gauge theory via the time-honored Weinberg sum rules (WSR)s \cite{Weinberg:1967kj}, in case of running or walking dynamics. By running dynamics here we mean that the coupling constant of the associated asymptotically free gauge theory has a dependence as function of energy similar to the one in Quantum Chromo Dynamics. In the {\it walking} dynamics regime the coupling constant is almost constant for a wide range of energy before resuming the running behavior at very high energies, higher than the scale below which chiral symmetry breaking occurs. 
In order to implement the walking dynamics we use and generalize the results presented
in \cite{Appelquist:1998xf}. This link allows us to provide important phenomenological relations for the
spectrum of the axial and vector type spin one mesons. 
We find that light vector mesons (around 1 TeV) are compatible only within the walking regime if one requires simultaneously that the
underlying strongly coupled theory leads to an $S$ parameter \cite{Peskin:1990zt} smaller than the one
associated to the first technicolor models (which were based on an SU(3) gauge theory with two Dirac techniflavors \cite{TC}). We also show how naturally the walking dynamics merges into the running one.

Our analysis and effective Lagrangian are sufficiently general to be applicable to the vast majority of models of dynamical electroweak symmetry breaking in agreement with the precision tests, and for which a strongly coupled four dimensional gauge theory underlies the dynamics.  Since the global symmetry is SU(4) for the MWT, one can easily reduce our model to the case of an SU(2)$_{\rm L}\times$SU(2)$_{\rm R}$ chiral symmetry.  For certain values of the couplings of the low energy effective theory - before imposing the generalized WSRs - one recovers the BESS models \cite{Casalbuoni:1995qt}. 

Walking dynamics differs from the QCD-like running behavior because of a nearby infrared (IR) fixed point which dominates the low energy dynamics. The physics of the fixed point theory per se is very interesting. If one assumes the existence of a theory with an actual IR fixed point coupled to a non-conformal theory (such as the standard model), in the way described recently by Georgi \cite{Georgi:2007ek,Georgi:2007si}, this leads to interesting phenomenology \cite{Cheung:2007ue,Greiner:2007hr,Davoudiasl:2007jr,Chen:2007je}. The presence of a conformal symmetry signals itself in a way that {\it formally} resembles the production of a non-integer number of massless invisible particles, where the non-integer number is nothing but the scale dimension of the conformal-sector opertor, which is weakly coupled to the standard model operators.  We expect, however, following reference \cite{Fox:2007sy}, that the coupling with the standard model fields will push the {\it unparticle} sector away from the IR fixed point.  If this is the case, in practice one will observe a walking dynamics in certain sectors of the theory, such as for example the electroweak symmetry breaking sector.  Our model should then be a reasonable description of a near conformal dynamics associated to this sector.

The rather comprehensive model we are going to develop in the following sections has been conceived in a way to ease its implementation on computer programs aiming to  provide interesting experimental signals for the physics at colliders.

\section{The Underlying Lagrangian for Minimal Walking Technicolor}

The new dynamical sector we consider, which underlies the Higgs mechanism, is an SU(2) technicolor gauge theory with two adjoint
technifermions \cite{Sannino:2004qp}. The theory is asymptotically free if the number of flavors $N_f$ is less than $2.75$.

The two adjoint fermions may be written as \beq Q_L^a=\left(\begin{array}{c} U^{a} \\D^{a} \end{array}\right)_L , \qquad U_R^a \
, \quad D_R^a \ ,  \qquad a=1,2,3 \ ,\eeq with $a$ being the adjoint color index of SU(2). The left handed fields are arranged in three
doublets of the SU(2)$_L$ weak interactions in the standard fashion. The condensate is $\langle \bar{U}U + \bar{D}D \rangle$ which
correctly breaks the electroweak symmetry.

The model as described so far suffers from the Witten topological anomaly \cite{Witten:fp}. However, this can easily be solved by
adding a new weakly charged fermionic doublet which is a technicolor singlet \cite{Dietrich:2005jn}. Schematically: 
\beq L_L =
\left(
\begin{array}{c} N \\ E \end{array} \right)_L , \qquad N_R \ ,~E_R \
. \eeq In general, the gauge anomalies cancel using the following
generic hypercharge assignment
\begin{align}
Y(Q_L)=&\frac{y}{2} \ ,&\qquad Y(U_R,D_R)&=\left(\frac{y+1}{2},\frac{y-1}{2}\right) \ , \label{assign1} \\
Y(L_L)=& -3\frac{y}{2} \ ,&\qquad
Y(N_R,E_R)&=\left(\frac{-3y+1}{2},\frac{-3y-1}{2}\right) \ \label{assign2} ,
\end{align}
where the parameter $y$ can take any real value \cite{Dietrich:2005jn}. In our notation
the electric charge is $Q=T_3 + Y$, where $T_3$ is the weak
isospin generator. One recovers the SM hypercharge
assignment for $y=1/3$.

To discuss the symmetry properties of the
theory it is
convenient to use the Weyl basis for the fermions and arrange them in the following vector transforming according to the
fundamental representation of SU(4)
\beq Q= \begin{pmatrix}
U_L \\
D_L \\
-i\sigma^2 U_R^* \\
-i\sigma^2 D_R^*
\end{pmatrix},
\label{SU(4)multiplet} \eeq where $U_L$ and $D_L$ are the left
handed techniup and technidown, respectively and $U_R$ and $D_R$ are
the corresponding right handed particles. Assuming the standard
breaking to the maximal diagonal subgroup, the SU(4) symmetry
spontaneously breaks to $SO(4)$. Such a breaking is driven by the
following condensate \beq \langle Q_i^\alpha Q_j^\beta
\epsilon_{\alpha \beta} E^{ij} \rangle =-2\langle \overline{U}_R U_L
+ \overline{D}_R D_L\rangle \ , \label{conde}
 \eeq
where the indices $i,j=1,\ldots,4$ denote the components
of the tetraplet of $Q$, and the Greek indices indicate the ordinary
spin. The matrix $E$ is a $4\times 4$ matrix defined in terms
of the 2-dimensional unit matrix as
 \beq E=\left(
\begin{array}{cc}
0 & \mathbbm{1} \\
\mathbbm{1} & 0
\end{array}
\right) \ . \eeq

We follow the notation of Wess and Bagger \cite{WessBagger}
$\epsilon_{\alpha \beta}=-i\sigma_{\alpha\beta}^2$ and $\langle
 U_L^{\alpha} {{U_R}^{\ast}}^{\beta} \epsilon_{\alpha\beta} \rangle=
 -\langle  \overline{U}_R U_L
 \rangle$. A similar expression holds for the $D$ techniquark.
The above condensate is invariant under an $SO(4)$ symmetry. This leaves us with nine broken  generators with associated Goldstone bosons.

Replacing the Higgs sector of the SM with the MWT the Lagrangian now
reads:
\begin{eqnarray}
\mathcal{L}_H &\rightarrow &  -\frac{1}{4}{\cal F}_{\mu\nu}^a {\cal F}^{a\mu\nu} + i\bar{Q}_L
\gamma^{\mu}D_{\mu}Q_L + i\bar{U}_R \gamma^{\mu}D_{\mu}U_R +
i\bar{D}_R \gamma^{\mu}D_{\mu}D_R \nonumber \\
&& +i \bar{L}_L \gamma^{\mu} D_{\mu} {L}_L +
i\bar{N}_R \gamma^{\mu}D_{\mu}N_R + i\bar{E}_R
\gamma^{\mu}D_{\mu}E_R
\end{eqnarray}
with the technicolor field strength ${\cal F}_{\mu\nu}^a =
\partial_{\mu}{\cal A}_{\nu}^a - \partial_{\nu}{\cal A}_{\mu}^a + g_{TC} \epsilon^{abc} {\cal A}_{\mu}^b
{\cal A}_{\nu}^c,\ a,b,c=1,\ldots,3$.
For the left handed techniquarks the covariant derivative is:

\begin{eqnarray}
D_{\mu} Q^a_L &=& \left(\delta^{ac}\partial_{\mu} + g_{TC}{\cal
A}_{\mu}^b \epsilon^{abc} - i\frac{g}{2} \vec{W}_{\mu}\cdot
\vec{\tau}\delta^{ac} -i g'\frac{y}{2} B_{\mu} \delta^{ac}\right)
Q_L^c \ .
\end{eqnarray}
${\cal A}_{\mu}$ are the techni gauge bosons, $W_{\mu}$ are the
gauge bosons associated to SU(2)$_L$ and $B_{\mu}$ is the gauge
boson associated to the hypercharge. $\tau^a$ are the Pauli matrices
and $\epsilon^{abc}$ is the fully antisymmetric symbol. In the case
of right handed techniquarks the third term containing the weak
interactions disappears and the hypercharge $y/2$ has to be replaced
according to whether it is an up or down techniquark. For the
left-handed leptons the second term containing the technicolor
interactions disappears and $y/2$ changes to $-3y/2$. Only the last
term is present for the right handed leptons with an appropriate
hypercharge assignment.

\section{Low Energy Theory for MWT}

We construct the effective theory for MWT including composite scalars and vector bosons, their self interactions, and their interactions with the electroweak gauge fields and the standard model fermions

\subsection{Scalar Sector} \label{sec:scalar}
The relevant effective theory for the Higgs sector at the electroweak scale consists, in our model, of a composite Higgs and its pseudoscalar partner, as well as nine pseudoscalar Goldstone bosons and their scalar partners. These
can be assembled in the matrix
\begin{eqnarray}
M = \left[\frac{\sigma+i{\Theta}}{2} + \sqrt{2}(i\Pi^a+\widetilde{\Pi}^a)\,X^a\right]E \ ,
\label{M}
\end{eqnarray}
which transforms under the full SU(4) group according to
\begin{eqnarray}
M\rightarrow uMu^T \ , \qquad {\rm with} \qquad u\in {\rm SU(4)} \ .
\end{eqnarray}
The $X^a$'s, $a=1,\ldots,9$ are the generators of the SU(4) group which do not leave  the vacuum expectation value (VEV) of $M$ invariant
\begin{eqnarray}
\langle M \rangle = \frac{v}{2}E
 \ .
\end{eqnarray}
Note that the notation used is such that $\sigma$ is a \emph{scalar}
while the $\Pi^a$'s are \emph{pseudoscalars}. It is convenient to
separate the fifteen generators of SU(4) into the six that leave the
vacuum invariant, $S^a$, and the remaining nine that do not, $X^a$.
Then the $S^a$ generators of the SO(4) subgroup satisfy the relation
\begin{eqnarray}
S^a\,E + E\,{S^a}^{T} = 0 \ ,\qquad {\rm with}\qquad  a=1,\ldots  ,  6 \ ,
\end{eqnarray}
so that $uEu^T=E$, for $u\in$ SO(4). The explicit realization of the generators is shown in appendix \ref{appgen}.

Notice that it is necessary to introduce the ``tilde'' fields in the matrix $M$ when realizing the global symmetry linearly. In fact, it can easily be shown that the matrix
\begin{eqnarray}
M = \left(\frac{\sigma}{2} + i\sqrt{2}\Pi^a\,X^a\right)E \nonumber
\end{eqnarray}
is not invariant in form under a general SU(4) transformation, but only under transformations of the unbroken SO(4) subgroup. This is
in contrast to the case of an ${\rm SU(2)}_{\rm L}\times {\rm SU(2)}_{\rm R}$ chiral group, whose minimal form involves a scalar
Higgs and three pseudoscalar Goldstone bosons only, but is similar to the case of an ${\rm SU(3)}_{\rm L}\times {\rm SU(3)}_{\rm R}$
chiral group. With the tilde fields included, the matrix $M$ is invariant in form under U(4)$\equiv$SU(4)$\times$U(1)$_{\rm
A}$, rather than just SU(4). However the U(1)$_{\rm A}$ axial symmetry is anomalous, and is therefore broken at the quantum level.

The connection between the composite scalars and the underlying techniquarks can be derived from the transformation properties under SU(4), by observing that the elements of the matrix $M$ transform like techniquark bilinears:
\begin{eqnarray}
M_{ij} \sim Q_i^\alpha Q_j^\beta \varepsilon_{\alpha\beta} \quad\quad\quad {\rm with}\ i,j=1\dots 4.
\label{M-composite}
\end{eqnarray}
Using this expression, and the basis matrices given in appendix \ref{appgen}, the scalar fields can be related to the wavefunctions of the techniquark bound states. This gives the following charge eigenstates:
\begin{eqnarray}
\begin{array}{rclcrcl}
v+H & \equiv & \sigma \sim  \overline{U}U+\overline{D}D  &,~~~~ &
\Theta  &\sim& i \left(\overline{U} \gamma^5 U+\overline{D} \gamma^5 D\right) \ ,  \\
A^0 & \equiv & \widetilde{\Pi}^3  \sim  \overline{U}U-\overline{D}D &,~~~~ &
\Pi^0 & \equiv & \Pi^3 \sim i \left(\overline{U} \gamma^5 U-\overline{D} \gamma^5 D\right) \ , \\
A^+ & \equiv & {\displaystyle \frac{\widetilde{\Pi}^1 - i \widetilde{\Pi}^2}{\sqrt{2}}} \sim \overline{D}U &,~~~~&
\Pi^+ & \equiv & {\displaystyle \frac{\Pi^1 - i \Pi^2}{\sqrt{2}}} \sim i \overline{D} \gamma^5 U \ , \\
A^- & \equiv & {\displaystyle \frac{\widetilde{\Pi}^1 + i \widetilde{\Pi}^2}{\sqrt{2}}} \sim \overline{U}D &,~~~~&
\Pi^- & \equiv & {\displaystyle \frac{\Pi^1 + i \Pi^2}{\sqrt{2}}} \sim i \overline{U} \gamma^5 D \ ,
\end{array}
\label{TM-eigenstates}
\end{eqnarray}
for the technimesons, and
\begin{eqnarray}
\begin{array}{rcl}
\Pi_{UU} & \equiv & {\displaystyle \frac{\Pi^4 + i \Pi^5 + \Pi^6 + i \Pi^7}{2}} \sim U^T C U \ , \\
\Pi_{DD} & \equiv & {\displaystyle \frac{\Pi^4 + i \Pi^5 - \Pi^6 - i \Pi^7}{2}} \sim D^T C D \ , \\
\Pi_{UD} & \equiv & {\displaystyle \frac{\Pi^8 + i \Pi^9}{\sqrt{2}}} \sim U^T C D \ , \\
\widetilde{\Pi}_{UU} & \equiv &
{\displaystyle \frac{\widetilde{\Pi}^4 + i \widetilde{\Pi}^5 + \widetilde{\Pi}^6 + i \widetilde{\Pi}^7}{2}} \sim i U^T C \gamma^5 U \ , \\
\widetilde{\Pi}_{DD} & \equiv &
{\displaystyle \frac{\widetilde{\Pi}^4 + i \widetilde{\Pi}^5 - \widetilde{\Pi}^6 - i \widetilde{\Pi}^7}{2}} \sim i D^T C \gamma^5 D  \ , \\
\widetilde{\Pi}_{UD} & \equiv & {\displaystyle \frac{\widetilde{\Pi}^8 + i \widetilde{\Pi}^9}{\sqrt{2}}} \sim i U^T C \gamma^5 D \ ,
\end{array}
\label{TB-eigenstates}
\end{eqnarray}
for the technibaryons, where $U\equiv (U_L,D_L)^T$ and $D\equiv (D_L,D_R)^T$ are Dirac technifermions, and $C$ is the charge conjugation matrix, needed to form Lorentz-invariant objects. To these technibaryon charge eigenstates we must add the corresponding charge conjugate states ({\em e.g.} $\Pi_{UU}\rightarrow \Pi_{\overline{U}\overline{U}}$).

The electroweak subgroup can be embedded in SU(4), as explained in detail in \cite{Appelquist:1999dq}. Here SO(4) acts as a vectorial subgroup, in the sense that this is the diagonal subgroup to which SU(4) is maximally broken. Based on this, we can say that the generators $S^a$, with $a=1,2,3$, form a vectorial SU(2) subgroup of SU(4), which is henceforth denoted by SU(2)$_{\rm V}$, while $S^4$ forms a U(1)$_{\rm V}$ subgroup. The $S^a$ generators, with $a=1,..,4$, together with the $X^a$ generators, with $a=1,2,3$, generate an SU(2)$_{\rm L}\times$SU(2)$_{\rm R}\times$U(1)$_{\rm V}$ algebra. This is easily seen by changing genarator basis from $(S^a,X^a)$ to $(L^a,R^a)$, where
\begin{eqnarray}
L^a \equiv \frac{S^a + X^a}{\sqrt{2}} = \begin{pmatrix}\frac{\tau^a}{2}\ \ \  & 0 \\ 0 & 0\end{pmatrix} \ , \ \
{-R^a}^T \equiv \frac{S^a-X^a}{\sqrt{2}}  = \begin{pmatrix}0 & 0 \\ 0 & -\frac{{\tau^a}^T}{2}\end{pmatrix} \ ,
\end{eqnarray}
with $a=1,2,3$. The electroweak gauge group is then obtained by gauging ${\rm SU(2)}_{\rm L}$, and the ${\rm U(1)}_Y$ subgroup of ${\rm SU(2)}_{\rm R}\times {\rm U(1)}_{\rm V}$, where
\begin{eqnarray}
Y =  -{R^3}^T + \sqrt{2}\ Y_{\rm V}\ S^4 \ ,
\end{eqnarray}
and $Y_{\rm V}$ is the U(1)$_{\rm V}$ charge. For example, from Eq.~(\ref{assign1}) and Eq.~(\ref{assign2}) we see that $Y_{\rm V}=y$ for the techniquarks, and $Y_{\rm V}=-3y$ for the new leptons. As SU(4) spontaneously breaks to SO(4), ${\rm SU(2)}_{\rm L}\times {\rm SU(2)}_{\rm R}$ breaks to ${\rm SU(2)}_{\rm V}$. As a consequence, the electroweak symmetry breaks to ${\rm U(1)}_Q$, where
\begin{eqnarray}
Q = \sqrt{2}\ S^3 + \sqrt{2}\ Y_{\rm V} \ S^4 \ .
\end{eqnarray}
Moreover the ${\rm SU(2)}_{\rm V}$ group, being entirely contained in the unbroken SO(4), acts as a custodial isospin, which insures that the $\rho$ parameter is equal to one at tree-level.

The electroweak covariant derivative for the $M$ matrix is
\begin{eqnarray}
D_{\mu}M =\partial_{\mu}M - i\,g \left[G_{\mu}(y)M + MG_{\mu}^T(y)\right]  \
, \label{covariantderivative}
\end{eqnarray}
where
\begin{eqnarray}
g\ G_{\mu}(Y_{\rm V}) & = & g\ W^a_\mu \ L^a + g^{\prime}\ B_\mu \ Y  \nonumber \\
& = & g\ W^a_\mu \ L^a + g^{\prime}\ B_\mu \left(-{R^3}^T+\sqrt{2}\ Y_{\rm V}\ S^4\right) \ .
\label{gaugefields}
\end{eqnarray}
Notice that in the last equation $G_\mu(Y_{\rm V})$ is written for a general U(1)$_{\rm V}$ charge $Y_{\rm V}$, while in Eq.~(\ref{covariantderivative}) we have to take the U(1)$_{\rm V}$ charge of the techniquarks, $Y_{\rm V}=y$, since these are the constituents of the matrix $M$, as explicitly shown in Eq.~(\ref{M-composite}).

Three of the nine Goldstone bosons associated with the broken generators become the longitudinal degrees of freedom of
the massive weak gauge bosons, while the extra six Goldstone bosons will acquire a mass due to extended technicolor interactions (ETC) as well as the
electroweak interactions per se. Using a bottom up approach we will not commit to a specific ETC theory but limit ourself to introduce the minimal low energy operators  needed to construct a phenomenologically viable theory. The new Higgs Lagrangian is
\begin{eqnarray}
{\cal L}_{\rm Higgs} &=& \frac{1}{2}{\rm Tr}\left[D_{\mu}M D^{\mu}M^{\dagger}\right] - {\cal V}(M) + {\cal L}_{\rm ETC} \ ,
\end{eqnarray}
where the potential reads
\begin{eqnarray}
{\cal V}(M) & = & - \frac{m^2}{2}{\rm Tr}[MM^{\dagger}] +\frac{\lambda}{4} {\rm Tr}\left[MM^{\dagger} \right]^2 
+ \lambda^\prime {\rm Tr}\left[M M^{\dagger} M M^{\dagger}\right] \nonumber \\
& - & 2\lambda^{\prime\prime} \left[{\rm Det}(M) + {\rm Det}(M^\dagger)\right] \ ,
\end{eqnarray}
and ${\cal L}_{\rm ETC}$ contains all terms which are generated by the ETC interactions, and not by the chiral symmetry breaking sector. Notice that the determinant terms (which are renormalizable) explicitly break the U(1)$_{\rm A}$ symmetry, and give mass to $\Theta$, which would otherwise be a massless Goldstone boson. While the potential has a (spontaneously broken) SU(4) global symmetry, the largest global symmetry of the kinetic term is SU(2)$_{\rm L}\times$U(1)$_{\rm R}\times$U(1)$_{\rm V}$ (where U(1)$_{\rm R}$ is the $\tau^3$ part of SU(2)$_{\rm R}$), and becomes SU(4) in the $g,g^\prime\rightarrow 0$ limit. Under electroweak gauge transformations, $M$ transforms like
\begin{eqnarray}
M(x) \rightarrow u(x;y) \ M(x) \ u^T(x;y) \ ,
\label{transf-M}
\end{eqnarray}
where
\begin{eqnarray}
u(x;Y_{\rm V}) = \exp{\left[i\alpha^a(x)L^a+i\beta(x)\left(-{R^3}^T+\sqrt{2}\ Y_{\rm V}\ S^4\right)\right]} \ ,
\label{u}
\end{eqnarray}
and $Y_{\rm V}=y$.
We explicitly break the SU(4) symmetry in order to provide mass to the Goldstone bosons which are not eaten by the weak gauge bosons. We, however, preserve the full 
SU(2)$_{\rm L}\times$SU(2)$_{\rm R}\times$U(1)$_{\rm V}$ subgroup of SU(4), since breaking SU(2)$_{\rm R}\times$U(1)$_{\rm V}$ to U(1)$_Y$ would result in a potentially dangerous violation of the custodial isospin symmetry. Assuming parity invariance we write: 
\begin{eqnarray}
{\cal L}_{\rm ETC} = \frac{m_{\rm ETC}^2}{4}\ {\rm Tr}\left[M B M^\dagger B + M M^\dagger \right] + \cdots \ ,
\end{eqnarray}
where the ellipses represent possible higher dimensional operators, and $B\equiv 2\sqrt{2}S^4$ commutes with the SU(2)$_{\rm L}\times$SU(2)$_{\rm R}\times$U(1)$_{\rm V}$ generators.

The potential ${\cal V}(M)$ is SU(4) invariant. It produces a VEV
which parameterizes the techniquark condensate, and spontaneously
breaks SU(4) to SO(4). In terms of the model parameters the VEV is
\begin{eqnarray}
v^2=\langle \sigma \rangle^2 = \frac{m^2}{\lambda + \lambda^\prime - \lambda^{\prime\prime} } \ ,
\label{VEV}
\end{eqnarray}
while the Higgs mass is
\begin{eqnarray}
M_H^2 = 2\ m^2 \ .
\end{eqnarray}
The linear combination $\lambda + \lambda^{\prime} -
\lambda^{\prime\prime}$ corresponds to the Higgs self coupling in
the SM. The three pseudoscalar mesons $\Pi^\pm$, $\Pi^0$ correspond
to the three massless Goldstone bosons which are absorbed by the
longitudinal degrees of freedom of the $W^\pm$ and $Z$ boson. The
remaining six uneaten Goldstone bosons are technibaryons, and all
acquire tree-level degenerate mass through, not yet specified, ETC interactions:
\begin{eqnarray}
M_{\Pi_{UU}}^2 = M_{\Pi_{UD}}^2 = M_{\Pi_{DD}}^2 = m_{\rm ETC}^2  \ .
\end{eqnarray}
The remaining scalar and pseudoscalar masses are
\begin{eqnarray}
M_{\Theta}^2 & = & 4 v^2 \lambda^{\prime\prime} \nonumber \\
M_{A^\pm}^2 = M_{A^0}^2 & = & 2 v^2 \left(\lambda^{\prime}+\lambda^{\prime\prime}\right)
\end{eqnarray}
for the technimesons, and
\begin{eqnarray}
M_{\widetilde{\Pi}_{UU}}^2 = M_{\widetilde{\Pi}_{UD}}^2 = M_{\widetilde{\Pi}_{DD}}^2 =
m_{\rm ETC}^2 + 2 v^2 \left(\lambda^{\prime} + \lambda^{\prime\prime }\right) \ ,
\end{eqnarray}
for the technibaryons. 
To gain insight on some of the mass relations one can use \cite{Hong:2004td}.

\subsection{Vector Bosons}
The composite vector bosons of a theory with a global SU(4) symmetry are conveniently described by the four-dimensional traceless Hermitian matrix
\begin{eqnarray}
A^\mu = A^{a\mu} \ T^a \ ,
\end{eqnarray}
where $T^a$ are the SU(4) generators: $T^a=S^a$, for $a=1, \dots ,6$, and $T^{a+6}=X^a$, for $a=1, \dots ,9$. Under an arbitrary SU(4) transformation, $A^\mu$ transforms like
\begin{equation}
A^\mu \ \rightarrow \ u\ A^\mu \ u^\dagger \ ,\ \ \ {\rm where} \ u\in {\rm SU(4)} \ .
\label{vector-transform}
\end{equation}
Eq.~(\ref{vector-transform}), together with the tracelessness of the matrix $A_\mu$, gives the connection with the techniquark bilinears:
\begin{equation}
A^{\mu,j}_{i}  \sim \ Q^{\alpha}_i  \sigma^{\mu}_{\alpha \dot{\beta}}  \bar{Q}^{\dot{\beta},j}
- \frac{1}{4} \delta_{i}^j Q^{\alpha}_k  \sigma^{\mu}_{\alpha \dot{\beta}} \bar{Q}^{\dot{\beta},k} \ .
\end{equation}
Then we find the following relations between the charge eigenstates and the wavefunctions of the composite objects:
\begin{eqnarray}
\begin{array}{rclcrcl}
v^{0\mu} & \equiv & A^{3\mu} \sim \bar{U} \gamma^\mu U - \bar{D} \gamma^\mu D & , & 
a^{0\mu} & \equiv & A^{9\mu} \sim \bar{U} \gamma^\mu \gamma^5 U - \bar{D} \gamma^\mu \gamma^5 D \\
v^{+\mu} & \equiv & {\displaystyle \frac{A^{1\mu}-i A^{2\mu}}{\sqrt{2}}} \sim \bar{D} \gamma^\mu U & , &
a^{+\mu} & \equiv & {\displaystyle \frac{A^{7\mu}-i A^{8\mu}}{\sqrt{2}}} \sim  \bar{D} \gamma^\mu  \gamma^5 U \\
v^{-\mu} & \equiv & {\displaystyle \frac{A^{1\mu}+i A^{2\mu}}{\sqrt{2}}} \sim  \bar{U} \gamma^\mu D & , &  
a^{-\mu} & \equiv & {\displaystyle \frac{A^{7\mu}+i A^{8\mu}}{\sqrt{2}}} \sim  \bar{U} \gamma^\mu  \gamma^5 D \\
v^{4\mu} & \equiv & A^{4\mu} \sim \bar{U} \gamma^\mu U + \bar{D} \gamma^\mu D  & , & & &
\end{array}
\label{TMV-eigenstates}
\end{eqnarray}
for the vector mesons, and
\begin{eqnarray}
\begin{array}{rcl}
x_{UU}^\mu & \equiv & {\displaystyle \frac{A^{10\mu}+i A^{11\mu}+A^{12\mu}+ i A^{13\mu}}{2}} \sim   U^T C \gamma^\mu \gamma^5 U \ , \\
x_{DD}^\mu & \equiv & {\displaystyle \frac{A^{10\mu}+i A^{11\mu}-A^{12\mu}- i A^{13\mu}}{2}} \sim   D^T C \gamma^\mu \gamma^5 D \ , \\
x_{UD}^\mu & \equiv & {\displaystyle \frac{A^{14\mu}+i A^{15\mu}}{\sqrt{2}}} \sim  D^T C \gamma^\mu \gamma^5 U \ , \\
s_{UD}^\mu & \equiv & {\displaystyle \frac{A^{6\mu}-i A^{5\mu}}{\sqrt{2}}} \sim U^T C \gamma^\mu  D \ ,
\end{array}
\label{TBV-eigenstates}
\end{eqnarray}
for the vector baryons.

There are different approaches on how to introduce vector mesons at the effective Lagrangian level. At the tree level they are all equivalent. The main differences emerge when exploring quantum corrections.

In the appendix we will show how to introduce the vector mesons in a way that renders the following Lagrangian amenable to loop computations.  
Based on these premises, the kinetic Lagrangian is:
\begin{eqnarray}
{\cal L}_{\rm kinetic} = -\frac{1}{2}{\rm Tr}\Big[\widetilde{W}_{\mu\nu}\widetilde{W}^{\mu\nu}\Big] - \frac{1}{4}B_{\mu\nu}B^{\mu\nu}
-\frac{1}{2}{\rm Tr}\Big[F_{\mu\nu}F^{\mu\nu}\Big] + m_A^2 \ {\rm Tr}\Big[C_\mu C^\mu\Big] \ ,
\label{massterm}
\end{eqnarray}
where $\widetilde{W}_{\mu\nu}$ and $B_{\mu\nu}$ are the ordinary field strength tensors for the electroweak gauge fields. Strictly speaking the terms above are not only kinetic ones since the Lagrangian contains a mass term as well as self interactions. The tilde on $W^a$ indicates  that the associated states are not yet the standard model weak triplets: in fact these states mix with the composite vectors to form mass eigenstates corresponding to the ordinary $W$ and $Z$ bosons. $F_{\mu\nu}$ is the field strength tensor for the new SU(4) vector bosons,
\begin{eqnarray}
F_{\mu\nu} & = & \partial_\mu A_\nu - \partial_\nu A_\mu - i\tilde{g}\left[A_\mu,A_\nu\right]\ ,
\label{strength}
\end{eqnarray}
and the vector field $C_\mu$ is defined by
\begin{eqnarray}
C_\mu \ \equiv \ A_\mu \ - \ \frac{g}{\tilde{g}}\ G_\mu (y) \ .
\end{eqnarray}
As shown in the appendix this is the appropriate linear combination to take which transforms homogeneously under the electroweak symmetries:
\begin{eqnarray}
C_\mu(x) \ \rightarrow \ u(x;y)\ C_\mu(x) \ u(x;y)^\dagger \ ,
\label{transf-C}
\end{eqnarray}
where $u(x;Y_{\rm V})$ is given by Eq.~(\ref{u}). (Once again, the specific assignment $Y_{\rm V}=y$, due to the fact that the composite vectors are built out of techniquark bilinears.) The mass term in Eq.~(\ref{massterm}) is gauge invariant (see the appendix), and gives a degenerate mass to all composite vector bosons, while leaving the actual gauge bosons massless. (The latter acquire mass as usual from the covariant derivative term of the scalar matrix $M$, after spontaneous symmetry breaking.)

The $C_\mu$ fields couple with $M$ via gauge invariant operators. Up
to dimension four operators the Lagrangian is (see the appendix for a more general treatment):
\begin{eqnarray}
{\cal L}_{\rm M-C} & = & \tilde{g}^2\ r_1 \ {\rm Tr}\left[C_\mu C^\mu M M^\dagger\right]
+ \tilde{g}^2\ r_2 \ {\rm Tr}\left[C_\mu M {C^\mu}^T M^\dagger \right] \nonumber \\
& + & i \ \tilde{g}\ r_3 \ {\rm Tr}\left[C_\mu \left(M (D^\mu M)^\dagger - (D^\mu M) M^\dagger \right) \right]
+ \tilde{g}^2\ s \ {\rm Tr}\left[C_\mu C^\mu \right] {\rm Tr}\left[M M^\dagger \right] \ .
\end{eqnarray}
The dimensionless parameters $r_1$, $r_2$, $r_3$, $s$ parameterize
the strength of the interactions between the composite scalars and
vectors in units of $\tilde{g}$, and are therefore naturally
expected to be of order one. However, notice that for
$r_1=r_2=r_3=0$ the overall Lagrangian possesses two independent
SU(2)$_{\rm L}\times$U(1)$_{\rm R}\times$U(1)$_{\rm V}$ global
symmetries. One for the terms involving $M$ and one for the terms
involving $C_\mu$~\footnote{The gauge fields explicitly break the
original SU(4) global symmetry to SU(2)$_{\rm L}\times$U(1)$_{\rm
R}\times$ U(1)$_{\rm V}$, where U(1)$_{\rm R}$ is the $T^3$ part of
SU(2)$_{\rm R}$, in the SU(2)$_{\rm L}\times$SU(2)$_{\rm
R}\times$U(1)$_{\rm V}$ subgroup of SU(4).}. The Higgs potential
only breaks the symmetry associated with $M$, while leaving the
symmetry in the vector sector unbroken. This {\em enhanced symmetry}
guarantees that all $r$-terms are still zero after loop corrections.
Moreover if one chooses $r_1$, $r_2$, $r_3$ to be small the near enhanced symmetry will protect these values against large corrections \cite{Casalbuoni:1995qt,Appelquist:1999dq}.

We can also construct dimension four operators including only
$C_{\mu}$ fields. These new operators will not affect our analysis
but will be relevant when investigating corrections to the trilinear
and quadrilinear gauge bosons interactions. We will include these terms in appendix~\ref{app:lagr}.

\subsection{Fermions and Yukawa Interactions}
The fermionic content of the effective theory consists of the standard model quarks and leptons, the new lepton doublet $L=(N,E)$ introduced to cure the Witten anomaly, and a composite techniquark-technigluon doublet. 

We now consider the limit according to which the SU(4) symmetry is, at first, extended to ordinary quarks and leptons. Of course, we will need to break this symmetry to accommodate the standard model phenomenology. We start by arranging the SU(2) doublets in SU(4) multiplets as we did for the techniquarks in Eq.~(\ref{SU(4)multiplet}). We therefore introduce the four component vectors $q^i$ and $l^i$,
\begin{eqnarray} 
q^i= \begin{pmatrix}
u^i_L \\
d^i_L \\
-i\sigma^2 {u^i_R}^* \\
-i\sigma^2 {d^i_R}^*
\end{pmatrix}\ , \quad
l^i= \begin{pmatrix}
\nu^i_L \\
e^i_L \\
-i\sigma^2 {\nu^i_R}^* \\
-i\sigma^2 {e^i_R}^*
\end{pmatrix}\ ,
\end{eqnarray}
where $i$ is the generation index. Note that such an extended SU(4) symmetry automatically predicts the presence of a right handed neutrino for each generation. In addition to the standard model fields there is an SU(4) multiplet for the new leptons,
\begin{eqnarray}
L = \begin{pmatrix}
N_L \\
E_L \\
-i\sigma^2 {N_R}^* \\
-i\sigma^2 {E_R}^*
\end{pmatrix}\ ,
\end{eqnarray}
and a multiplet for the techniquark-technigluon bound state,
\begin{eqnarray} 
\widetilde{Q}= \begin{pmatrix}
\widetilde{U}_L \\
\widetilde{D}_L \\
-i\sigma^2 {\widetilde{U}_R}^* \\
-i\sigma^2 {\widetilde{D}_R}^*
\end{pmatrix}\ .
\end{eqnarray}
With this arrangement, the electroweak covariant derivative for the fermion fields can be written
\begin{eqnarray}
D_\mu \ = \  \partial_\mu \  - \  i \ g \ G_\mu (Y_{\rm V})  \ , 
\end{eqnarray}
where $Y_{\rm V}=1/3$ for the quarks, $Y_{\rm V}=-1$ for the leptons, $Y_{\rm V}=-3y$ for the new lepton doublet, and $Y_{\rm V}=y$ for the techniquark-technigluon bound state. One can check that these charge assignments give the correct electroweak quantum numbers for the standard model fermions. In addition to the covariant derivative terms, we should add a term coupling $\widetilde{Q}$ to the vector field $C_\mu$, which transforms globally under electroweak gauge transformations. Such a term naturally couples the composite fermions to the composite vector bosons which otherwise would only feel the week interactions. Based on this, we write the following gauge part of the fermion Lagrangian:
\begin{eqnarray}
{\cal L}_{\rm fermion} & = & i\ \overline{q}^i_{\dot{\alpha}}  \overline{\sigma}^{\mu,\dot{\alpha} \beta} D_\mu  q^i_\beta 
+ i\ \overline{l}^i_{\dot{\alpha}}  \overline{\sigma}^{\mu,\dot{\alpha} \beta} D_\mu  l^i_\beta 
+ i\ \overline{L}_{\dot{\alpha}}  \overline{\sigma}^{\mu,\dot{\alpha} \beta} D_\mu  L_\beta 
+ i\ \overline{\widetilde{Q}}_{\dot{\alpha}}  \overline{\sigma}^{\mu,\dot{\alpha} \beta} D_\mu  \widetilde{Q}_\beta  \nonumber \\
& + & x\ \overline{\widetilde{Q}}_{\dot{\alpha}}  \overline{\sigma}^{\mu,\dot{\alpha} \beta} C_\mu  \widetilde{Q}_\beta
\label{fermion-kinetic}
\end{eqnarray}
The terms coupling the standard model fermions or the new leptons to $C_\mu$ are in general not allowed. In fact under electroweak gauge transformations any four-component fermion doublet $\psi$ transforms like
\begin{eqnarray}
\psi \rightarrow  u(x;Y_{\rm V}) \ \psi \ ,
\label{transf-psi}
\end{eqnarray}
and from Eq.~(\ref{transf-C}) we see that a term like $\psi^\alpha  \sigma^{\mu}_{\alpha \dot{\beta}} C_\mu  \overline{\psi}^{\dot{\beta}}$ is only invariant if $Y_{\rm V}=y$. Then we can distinguish two cases. First, we can have
$y\neq 1/3$ and $y\neq -1$, in which case $\psi^\alpha  \sigma^{\mu}_{\alpha \dot{\beta}} C_\mu  \overline{\psi}^{\dot{\beta}}$ is only invariant for $\psi=\widetilde{Q}$. Interaction terms of the standard model fermions with components of $C_\mu$ are still possible, but these would break the SU(4) chiral simmetry even in the limit in which the electroweak gauge interactions are switched off. Second, we can have $y=1/3$ or $y=-1$. Then $\psi^\alpha  \sigma^{\mu}_{\alpha \dot{\beta}} C_\mu  \overline{\psi}^{\dot{\beta}}$ is not only invariant for $\psi=\widetilde{Q}$, but also for either $\psi=q^i$ or $\psi=l^i$, respectively. In the last two cases, however, the corresponding interactions are highly suppressed, since these give rise to anomalous couplings of the light fermions with the standard model gauge bosons, which are tightly constrained by experiments.

We now turn to the fundamental issue of providing masses to ordinary fermions.  Many  extensions of technicolor have been suggested in the literature to address this problem. Some of the extensions use another strongly coupled gauge dynamics,  others introduce fundamental scalars. Many variants of the schemes presented above exist and a review of the major models is the one by Hill and Simmons \cite{Hill:2002ap}. At the moment there is not yet a consensus on which is the correct ETC. To keep the number of fields minimal we make the most economical ansatz, i.e. we parameterize our ignorance about a complete ETC theory by simply coupling the fermions to our low energy effective Higgs. This simple construction minimizes the flavor changing neutral currents problem. It is worth mentioning that it is possible to engineer a schematic ETC model proposed first by Randall in \cite{Randall:1992vt} and adapted for the MWT in \cite{Evans:2005pu} for which the effective theory presented in the main text can be considered a minimal description. \footnote{Another non minimal way to give masses to the ordinary fermions is to (re)introduce a new Higgs doublet as already done many times in the literature \cite{Simmons:1988fu,Dine:1990jd,Kagan:1990az,Kagan:1991gh,Carone:1992rh,Carone:1993xc,Gudnason:2006mk}. This possibility and its phenomenological applications will be studied elsewhere.}

Depending on the value of $y$ for the techniquarks, we can write different Yukawa interactions which couple the standard model fermions to the matrix $M$. Let $\psi$ denote either $q^i$ or $l^i$. If $\psi$ and the techniquark multiplets $Q^a$ have the same U(1)$_{\rm V}$ charge, then the Yukawa term
\begin{eqnarray}
- \psi^T M^* \psi + {\rm h.c.} \ ,
\label{yukawa1}
\end{eqnarray}
is gauge invariant, as one can check explicitly from Eq.~(\ref{transf-M}) and Eq.~(\ref{transf-psi}). Otherwise, if $\psi$ and $Q^a$ have different U(1)$_{\rm V}$ charges, we can only write a gauge invariant Lagrangian with the off-diagonal terms of $M$, which contain the Higgs and the Goldstone bosons:
\begin{eqnarray}
- \psi^T M_{\rm off}^*\ \psi + {\rm h.c.} \ .
\label{yukawa2}
\end{eqnarray}
In fact $M_{\rm off}$ has no U(1)$_{\rm V}$ charge, since
\begin{eqnarray}
S^4 M_{\rm off} + M_{\rm off} {S^4}^T = 0 \ ,
\end{eqnarray}
The last equation implies that the U(1)$_{\rm V}$ charges of $\psi^T$ and $\psi$ cancel in Eq.~(\ref{yukawa2}). The latter is actually the only viable Yukawa Lagrangian for the new leptons, since the corresponding U(1)$_{\rm V}$ charge is $Y_{\rm V}=-3y \neq y$, and for the ordinary quarks, since Eq.~(\ref{yukawa1}) contains $qq$ terms which are not color singlets. 

We notice however that neither Eq.~(\ref{yukawa1}) nor Eq.~(\ref{yukawa2}) are phenomenologically viable yet, since they leave the SU(2)$_{\rm R}$ subgroup of SU(4) unbroken, and the corresponding Yukawa interactions do not distinguish between the up-type and the down-type fermions. In order to prevent this feature, and recover agreement with the experimental input, we break the SU(2)$_{\rm R}$ symmetry to U(1)$_{\rm R}$ by using the projection operators $P_U$ and $P_D$, where
\begin{eqnarray}
P_U = \begin{pmatrix} 1 & 0 \\ 0 & \frac{1+\tau^3}{2} \end{pmatrix} \ , \quad
P_D = \begin{pmatrix} 1 & 0 \\ 0 & \frac{1-\tau^3}{2} \end{pmatrix} \ .
\end{eqnarray}
Then, for example, Eq.~(\ref{yukawa1}) should be replaced by
\begin{eqnarray}
- \psi^T \left(P_U M^* P_U\right) \psi - \psi^T \left(P_D M^* P_D\right) \psi + {\rm h.c.} \ .
\label{yukawa3}
\end{eqnarray}

 {}For illustration we distinguish two different cases for our analysis, $y\neq -1$ and $y= -1$, and write the corresponding Yukawa interactions:
\newline
(i) $y\neq -1$. In this case we can only form gauge invariant terms with the standard model fermions by using the off-diagonal $M$ matrix. Allowing for both $N-E$ and $\widetilde{U}-\widetilde{D}$ mass splitting, we write
\begin{eqnarray}
{\cal L}_{\rm Yukawa} &=& -\ y_u^{ij}\ q^{i T} \left(P_U M_{\rm off}^* P_U\right) q^j
- y_d^{ij}\ q^{i T} \left(P_D M_{\rm off}^* P_D\right) q^j \nonumber \\
&& -\ y_\nu^{ij}\ l^{i T} \left(P_U M_{\rm off}^* P_U\right) l^j
- y_e^{ij}\ l^{i T} \left(P_D M_{\rm off}^* P_D\right) l^j \nonumber \\
&& -\ y_N\ L^T \left(P_U M_{\rm off}^* P_U\right) L
-\ y_E\ L^T \left(P_D M_{\rm off}^* P_D\right) L \nonumber \\
&& -\ \ y_{\widetilde{U}} \widetilde{Q}^T \left(P_U M^* P_U\right) \widetilde{Q} 
- \ y_{\widetilde{D}} \widetilde{Q}^T \left(P_D M^* P_D\right) \widetilde{Q} \ + {\rm h.c.} \ ,
\label{yukawa-1}
\end{eqnarray}
where $y_u^{ij}$, $y_d^{ij}$, $y_\nu^{ij}$, $y_e^{ij}$ are arbitrary complex matrices, and $y_N$, $y_E$, $y_{\widetilde{U}}$, $y_{\widetilde{D}}$ are complex numbers.
\newline
Note that the underlying strong dynamics already provides a dynamically generated mass term for $\widetilde{Q}$ of the type: 
\begin{equation}
{k}\,  \widetilde{Q}^T  M^* \widetilde{Q} + {\rm h.c.} \ ,
\end{equation}
with ${k}$ a dimensionless coefficient of order one and entirely fixed within the underlying theory.  The splitting between the up and down type techniquarks is due to physics beyond the technicolor interactions \footnote{Small splittings with respect to the electroweak scale will be induced by the standard model corrections per se.}. Hence the Yukawa interactions for $\widetilde{Q}$ must be interpreted as already containing the dynamical generated mass term.  

(ii) $y= -1$. In this case we can form gauge invariant terms with the standard model leptons and the full $M$ matrix:
\begin{eqnarray}
{\cal L}_{\rm Yukawa} &=& -\ y_u^{ij}\ q^{i T} \left(P_U M_{\rm off}^* P_U\right) q^j
- y_d^{ij}\ q^{i T} \left(P_D M_{\rm off}^* P_D\right) q^j \nonumber \\
&& -\ y_\nu^{ij}\ l^{i T} \left(P_U M^* P_U\right) l^j
- y_e^{ij}\ l^{i T} \left(P_D M^* P_D\right) l^j \nonumber \\
&& -\ y_N\ L^T \left(P_U M_{\rm off}^* P_U\right) L
-\ y_E\ L^T \left(P_D M_{\rm off}^* P_D\right) L \nonumber \\
&& -\ \ y_{\widetilde{U}} \widetilde{Q}^T \left(P_U M^* P_U\right) \widetilde{Q} 
- \ y_{\widetilde{D}} \widetilde{Q}^T \left(P_D M^* P_D\right) \widetilde{Q} \ + {\rm h.c.} \ .
\label{yukawa-2}
\end{eqnarray}
Here we are assuming Dirac masses for the neutrinos, but we can easily add also Majorana mass terms. At this point one can exploit the symmetries of the kinetic terms to induce a GIM mechanism, which works out exactly like in the standard model. Therefore, in both Eq.~(\ref{yukawa-1}) and Eq.~(\ref{yukawa-2}) we can assume $y_u^{ij}$, $y_d^{ij}$, $y_\nu^{ij}$, $y_e^{ij}$ to be diagonal matrices, and replace the $d^i_L$ and $\nu^i_L$ fields, in the kinetic terms, with $V_q^{ij} d^j_L$ and $V_l^{ij} \nu^j_L$, respectively, where $V_q$ and $V_l$ are the mixing matrices.  

When $y=-1$ $\widetilde{Q}$ has the same quantum numbers of the ordinary leptons, except for the technibaryon number. If the technibaryon number is violated they can mix with the ordinary leptons behaving effectively as a  fourth generation leptons (see Eq.~(\ref{yukawa-2})). However this will reintroduce, in general, anomalous couplings with intermediate gauge bosons for the ordinary fermions and hence we assume the mixing to be small.

\section{Weinberg Sum Rules and Electroweak Parameters}\label{sec:electroweak}

The effective theory described until now has a number of free
parameters which are fixed once the associated underlying dynamics is specified. Conversely, the low effective theory introduced above encompasses
different underlying realizations with the same symmetry pattern. In the following we partially reduce the arbitrariness of the Lagrangian by assuming that the underlying dynamcics is the one of a four dimensional asymptotically free gauge theory, with only fermionic matter fields transforming according to a given but otherwise arbitrary representation of the gauge group. The MWT is automatically part of this set of theories. We will use the Weinberg sum rules (WSR) as main ingredient to reduce the number of unknown parameters in the theory. We will then compute the universal corrections in the effective theory and compare these to the ones estimated in the underlying theory. The latter step is taken only for the MWT theory but it straightforwardly generalizes to any walking theory. We also use the results found in \cite{Appelquist:1998xf}  which allow us to treat walking and running theories in a unified way.

\subsection*{Weinberg sum rules}
Our effective theory is meant to be associated to an underlying strongly coupled theory. Hence we relate some of the parameters in the effective theory via the WSR. These are linked to the  two point vector-vector minus axial-axial vacuum polarization amplitude, which is known to be sensitive to chiral symmetry breaking.  We define
\begin{equation}
i\Pi_{\mu \nu}^{a,b}(q)\equiv \int\!d^4x\, e^{-i qx}
\left[<J_{\mu,V}^a(x)J_{\nu,V}^b(0)> -
 <J_{\mu,A}^a(x)J_{\nu,A}^b(0)>\right] \ ,
\label{VA}
\end{equation}
within the underlying strongly coupled gauge theory, where
\begin{equation}
\Pi_{\mu \nu}^{a,b}(q)=\left(q_{\mu}q_{\nu} - g_{\mu\nu}q^2 \right) \,
\delta^{a b} \Pi(q^2) \ .
\end{equation}
Here $a,b=1,...,N_f^2-1$, label the flavor
currents and the SU(N$_f$) generators are normalized according to
$\rm{Tr} \left[T^a T^b\right]= (1/2) \delta^{ab} $.  The
function $\Pi(q^2)$ obeys the unsubtracted dispersion relation
\begin{equation}
\frac{1}{\pi} \int_0^{\infty}\!ds\, \frac{{\rm Im}\Pi(s)}{s + Q^2}
=\Pi(Q^2) \ ,
\label{integral}
\end{equation}
where $Q^2=-q^2 >0$, and the constraint
$\displaystyle{-Q^2 \Pi(Q^2)>0}$ holds for $0 < Q^2 < \infty$~\cite{Witten:1983ut}. The discussion above is for the standard chiral symmetry breaking pattern SU(N$_f$)$\times$ SU(N$_f$)$ \rightarrow $SU(N$_f$) but it is generalizable to any breaking pattern.

Since we are imagining the underlying theory to be asymptotically
free, the behavior of $\Pi(Q^2)$ at asymptotically high momenta is
the same as in ordinary QCD, i.e. it scales like $Q^{-6}$~\cite{Bernard:1975cd}. Expanding the left hand side of the dispersion relation
thus leads to the two conventional spectral function sum rules
\begin{equation}
\frac{1}{\pi} \int_0^{\infty}\!ds\,{\rm Im}\Pi(s) =0
\label{spectral1}
\quad {\rm and} \quad
\frac{1}{\pi} \int_0^{\infty}\!ds\,s \,{\rm Im}\Pi(s) =0 \ .
\end{equation}
Walking dynamics affects only the second sum rule \cite{Appelquist:1998xf} which is more sensitive to large but not asymptotically large momenta due to fact that the associated integrand contains an extra power of $s$.

We now saturate the absorptive part of the vacuum
polarization. We follow reference \cite{Appelquist:1998xf} and hence divide the
energy range of integration in three parts. The resonance part
which we approximate by the vector and axial mesons as well as the
Goldstone bosons. The second one (the continuum region) where the walking
dynamics takes over and which extends up to the scale above which
the underlying coupling constant drops like in a QCD-like theory.

The first WSR implies:
\begin{equation}
F^2_V - F^2_A = F^2_{\pi}\ ,
\label{1rule}
\end{equation}
where $F^2_V$ and $F^2_A$ are the vector and axial mesons decay
constants.  This sum rule holds for walking and running dynamics. A
more general representation of the resonance spectrum would replace
the left hand side of this relation with a sum over vector and axial
states.

The second sum rule receives important contributions from throughout
the near conformal region and can be
expressed in the form of:
\begin{equation}
F^2_V M^2_V - F^2_A M^2_A = a\,\frac{8\pi^2}{d(R)}\,F_{\pi}^4,
\label{2rule-2}
\end{equation}
where $a$ is expected to be positive and $O(1)$ and $d(R)$ is the dimension of the representation of the underlying fermions.  As in  the
case of the first sum rule, a more general resonance spectrum will
lead to a left-hand side with a sum over vector and axial states. In
either case, the conformal region enhances the vector piece relative to
the axial. We have generalized the result of reference \cite{Appelquist:1998xf}  to the case in which the fermions belong to a generic representation of the gauge group. In the case of running dynamics the right-hand side of the previous equation vanishes.  

\subsection*{Relating WSRs to the Effective Theory \& $S$ parameter }
The $S$ parameter is related to the absorptive part  of the
vector-vector minus axial-axial vacuum polarization as follows:
\begin{equation}
S=4\int_0^\infty \frac{ds}{s} {\rm Im}\bar{\Pi}(s)= 4\pi
\left[\frac{F^2_V}{M^2_V} - \frac{F^2_A}{M^2_A} \right] \ ,
\label{s-def}
\end{equation}
where ${\rm Im}\bar{\Pi}$ is obtained from ${\rm Im}\Pi$ by
subtracting the Goldstone boson contribution.

Other attempts to estimate the $S$ parameter for walking technicolor
theories have been made in the past \cite{Sundrum:1991rf} showing reduction of the $S$ parameter. $S$ has also been evaluated using computations inspired by the original AdS/CFT correspondence \cite{Maldacena:1997re} in \cite{Hong:2006si,Hirn:2006nt,Piai:2006vz,Agashe:2007mc,Carone:2007md}. 

Very recently Kurachi, Shrock and Yamawaki \cite{Kurachi:2007at} have further confirmed the results presented in \cite{Appelquist:1998xf} with their computations tailored for describing four dimensional gauge theories near the conformal window.
The present approach \cite{Appelquist:1998xf} is more physical since it is based on the
nature of the spectrum of states associated directly to the underlying gauge theory.  

Note that in this work we are taking a rather conservative approach in which the $S$ parameter, although reduced with respect to the case of a running theory, is positive and not very small.  After all, other sectors of the theory such as new leptons can further reduce or even offset a positive value of $S$ due solely to the technicolor theory. 

In our effective theory the $S$ parameter is directly proportional to the parameter $r_3$ via:
\begin{eqnarray}
S= \frac{8\pi}{\tilde{g}^2}\, \chi \,(2- \chi)   \ , \quad {\rm with} \quad \chi = \frac{v^2\tilde{g}^2}{2M^2_A} \, r_3 \ ,
\label{S}
\end{eqnarray}
where we have expanded in $g/\tilde{g}$ and kept only the leading order. The full expression can be found in appendix D.
We can now use the  sum rules to relate $r_3$ to other parameters in the theory for the running and the walking case.  Within the effective theory we deduce:
\begin{eqnarray}
F^2_V = \left(1 - \, \chi \frac{r_2}{r_3}\right)\, \frac{2M^2_A}{\tilde{g}^2} = \frac{2M^2_V}{\tilde{g}^2}\ , \quad F^2_A = 2\frac{M^2_A}{\tilde{g}^2}(1 - \chi)^2 \ , \quad F^2_{\pi} = v^2 ( 1 -\chi \, r_3 ) \ .\end{eqnarray}
Hence the first WSR reads:
\begin{eqnarray}
1+r_2 - 2 r_3 = 0 \ ,
\end{eqnarray}
while the second:
\begin{eqnarray}
 (r_2 - r_3) (v^2\tilde{g}^2 (r_2 + r_3) - 4 M^2_A) = a \frac{16\pi^2}{d(R)}  v^2\left( 1 - \chi \, r_3\right)^2 \ .
\end{eqnarray}

To gain analytical insight we consider the limit in which $\tilde{g} $ is small while $g/\tilde{g}$ is still much smaller than one. To leading order in $\tilde{g}$ the second sum rule simplifies to:
\begin{eqnarray}
r_3 - r_2  = a \frac{4\pi^2}{d(R)}\frac{v^2}{M^2_A} \ ,
\end{eqnarray}
Together with the first sum rule we find:
\begin{eqnarray}
r_2 = 1 - 2 t \ , \qquad r_3 = 1 - t \ ,
\end{eqnarray}
with
\begin{eqnarray}
t= a \frac{4\pi^2}{d(R)}\frac{v^2}{M^2_A} \ .
\end{eqnarray}
The approximate $S$ parameter reads.
\begin{eqnarray}
S= 8\pi\, \frac{v^2}{M^2_A} (1-t)   \ . \end{eqnarray}
A small value of $a$ provides a large and positive $t$ rendering $S$ smaller than expected in a running theory. In the next subsection we will make a similar analysis without taking the limit of small $\tilde{g}$.

 \subsection*{Axial-Vector Spectrum via WSRs}

It is is interesting to determine the relative vector to axial
spectrum as function of one of the two masses, say the axial one,
for a fixed value of the $S$ parameter meant to be associated to a given underlying gauge theory. 

\begin{figure}[!t]
\centering
\resizebox{10cm}{!}{\includegraphics{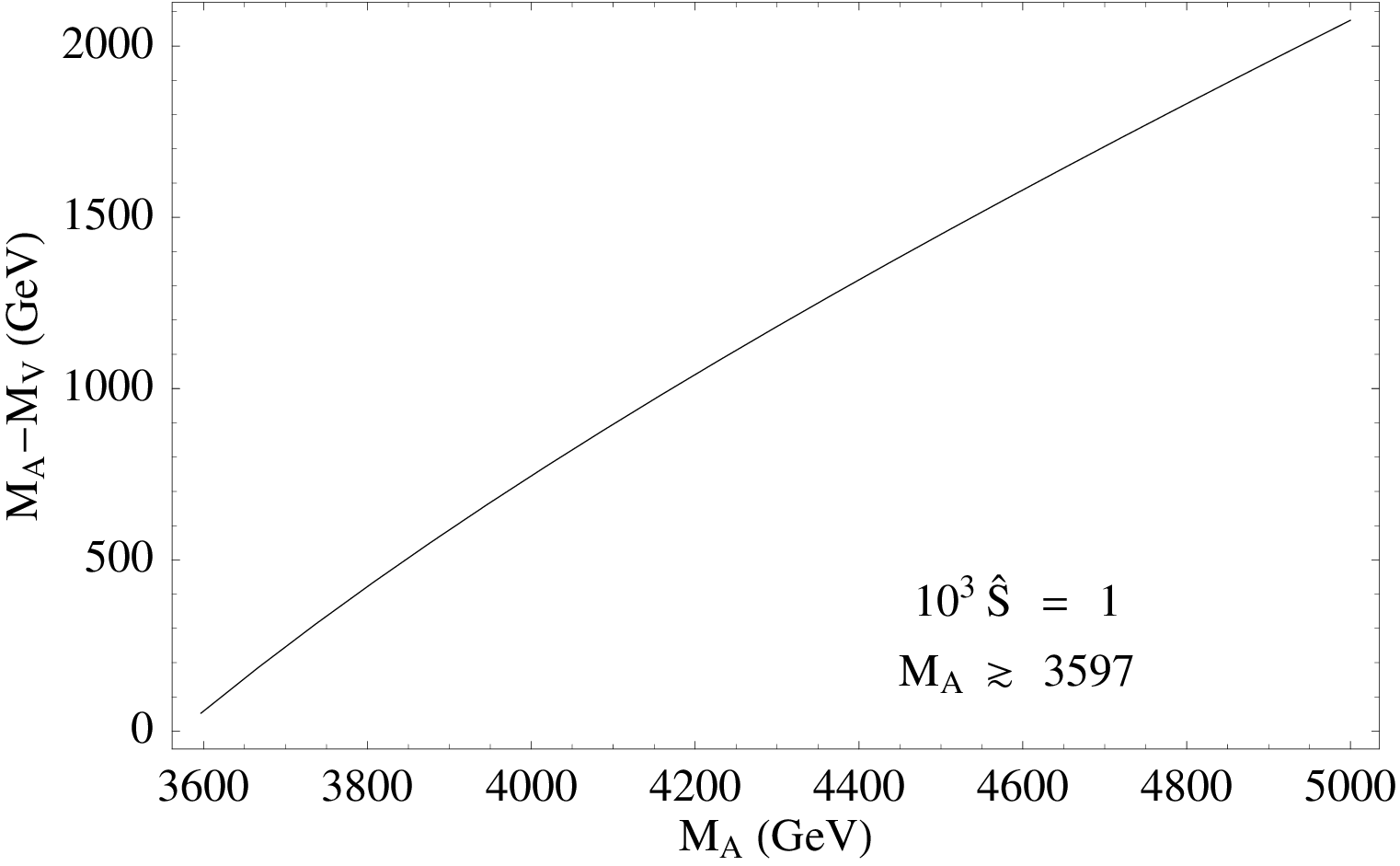}}
\caption{In the picture above we  have set $10^3 \hat{S}=1$, corresponding to $S\simeq 0.11$.  In the appendix we have provided the relation between $\hat{S}$ and the traditional S. Here we have imposed the first and the second WSR for $a=0$. This corresponds to an underlying gauge theory with a standard running behavior of the coupling constant. }
\label{runningwsr}
\end{figure}

For a running type dynamics (i.e. $a=0$) the two WSRs force the vector mesons to be quite heavy (above 3 TeV) in order to have a relatively low $S$ parameter ($S\simeq$ 0.1). This can be seen directly from Eq.~(\ref{S}) in the running regime, where $r_2=r_3=1$. This leads to
\begin{eqnarray}
M_A^2 \ \gtrsim \ \frac{8\pi v^2}{S} \ ,
\label{lowMA}
\end{eqnarray}
which corresponds to $M_A\gtrsim$ 3.6 TeV, for $S\simeq$ 0.11. Perhaps a more physical way to express this is to say that it is hard to have an intrinsically small $S$ parameter for running type theories. By small we mean smaller than the scaled up technicolor version of QCD with two techniflavors, in which $S\simeq$ 0.3. In Figure \ref{runningwsr} we plot the difference between the axial and vector mass as function of the axial mass, for $S\simeq$0.11. Since Eq.~(\ref{lowMA}) provides a lower bound for $M_A$, this plot shows that in the running regime the axial mass is always heavier than the vector mass. In fact the $M_A^2-M_V^2$ difference is proportional to $r_2$, with a positive proportionality factor (see the appendix), and $r_2=1$ in the running regime.

\begin{figure}[!t]
\centering
\begin{tabular}{cc}
\hskip -1.2cm\resizebox{8.5cm}{!}{\includegraphics{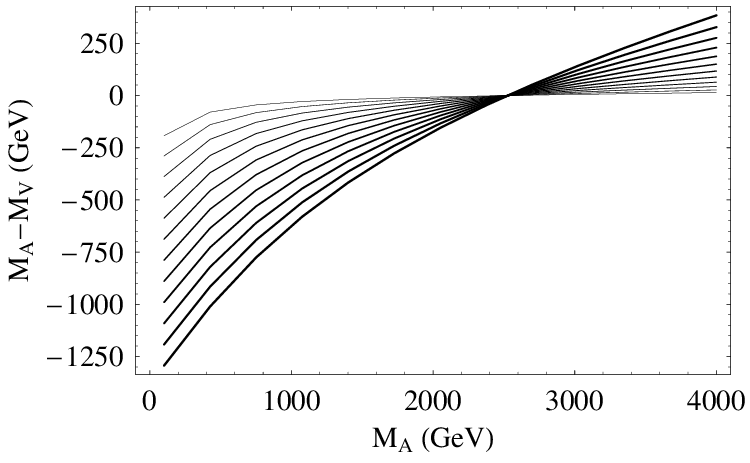}} & \resizebox{8.5cm}{!}{\includegraphics{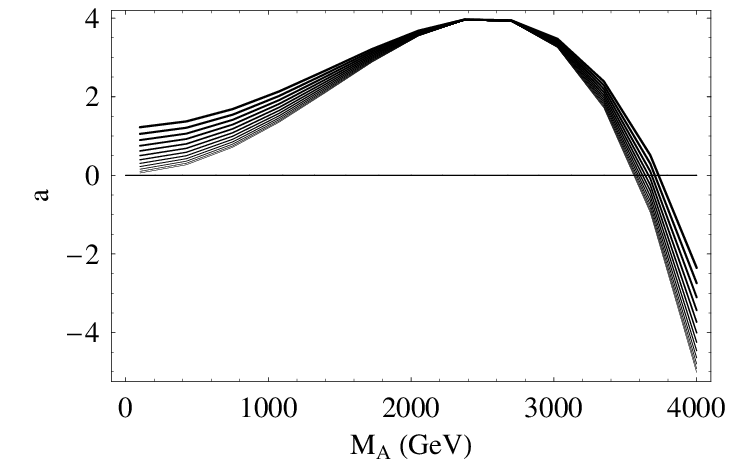}}
\end{tabular}
\caption{In the two pictures above we have set $10^3 \hat{S}=1$, corresponding to $S\simeq 0.11$, and the different curves are obtained by varying $\tilde{g}$ from one (the thinnest curve)  to eight  (the thickest curve). We have imposed the first WSR. {\it Left Panel}: We plot the allowed values of $M_A-M_V$ as function of $M_A$. {\it Right Panel}: We compute the value that $a$ should assume as function of $M_A$ in order for the second WSR to be satisfied in the walking regime. Note that $a$ is expected to be positive or zero. }
\label{walkwsr}
\end{figure}

When considering the second WSR modified by the walking dynamics, we observe that it is possible to have quite light spin one  vector mesons compatible with a small $S$ parameter. We numerically solve the first and second WSR in presence of the contribution due to walking in the second sum rule. The results are summarized in Figure \ref{walkwsr}. As for the running case we set again $S\simeq 0.11$. This value is close to the estimate in the underlying MWT \footnote{For the MWT we separate the contribution due to the new leptonic sector (which will be dealt with later in the main text) and the one due to the underlying strongly coupled gauge theory which is expected to be well represented by the perturbative contribution and is of the order of $1/2\pi$. When comparing with the $S$ parameter from the vector meson sector of the effective theory we should subtract from the underlying $S$ the one due to the new fermionic composite states $\tilde{U}$ and $\tilde{D}$. This contribution is very small since it is $1/6\pi$ in the limit when these states are degenerate and heavier than the zed gauge boson.}. The different curves are obtained by varying $\tilde{g}$ from one (the thinnest curve)  to eight (the thickest curve). We plot the allowed values of $M_A-M_V$ as function of $M_A$ in the left panel, having imposed only the first sum rule. In the right panel we compute the corresponding value that $a$ should assume as function of $M_A$ in order for the second WSR to be satisfied in the walking regime as well.
For any given underlying gauge theory all of the values of the parameters are fixed and our computation shows that it is possible to have walking theories with light vector mesons and a small $S$ parameter. Such a scenario needs a positive value of $a$, together with an axial meson lighter than its associated vector meson for $a$ between zero and four, when the axial vector mass is lighter than approximatively 2.5 TeV. However for spin one fields heavier than roughly 2.5 TeV and with still a positive $a$ one has an axial meson heavier than the vector one.  A degenerate spectrum allows for a small $S$ but with  relatively large values of $a$ and spin one masses around 2.5 TeV. We observe that $a$ becomes zero (and eventually negative) when the vector spectrum becomes sufficiently heavy. In other words we recover the running behavior for large masses of spin-one fields. Although in the plot we show negative values of $a$ one should stop the analysis after having reached a zero value of $a$. In fact, for masses heavier than roughly 3.5 TeV the second WSR for the running behavior, i.e. $a=0$, is enforced.  

Our results are general and help elucidating how different underlying dynamics will manifest itself at LHC. Any four dimensional strongly interacting theory replacing the Higgs mechanism, with two Dirac techniflavors gauged under the electroweak theory, is expected to have a spectrum of the low lying vector resonances like the one presented above.

We provide the relation between $S$ and the recently proposed modification of the electroweak parameters \cite{Barbieri:2004qk} in appendix \ref{electrocorrections}. In the same appendix we explicitly compute the remaining parameters within our effective theory and check that they do not lead to further constraints given the current status of the precision measurements.

\subsection*{Vanishing $S$ via New Leptons}
\begin{figure}[!t]
\centering
\begin{tabular}{cc}
\resizebox{6.5cm}{!}{\includegraphics{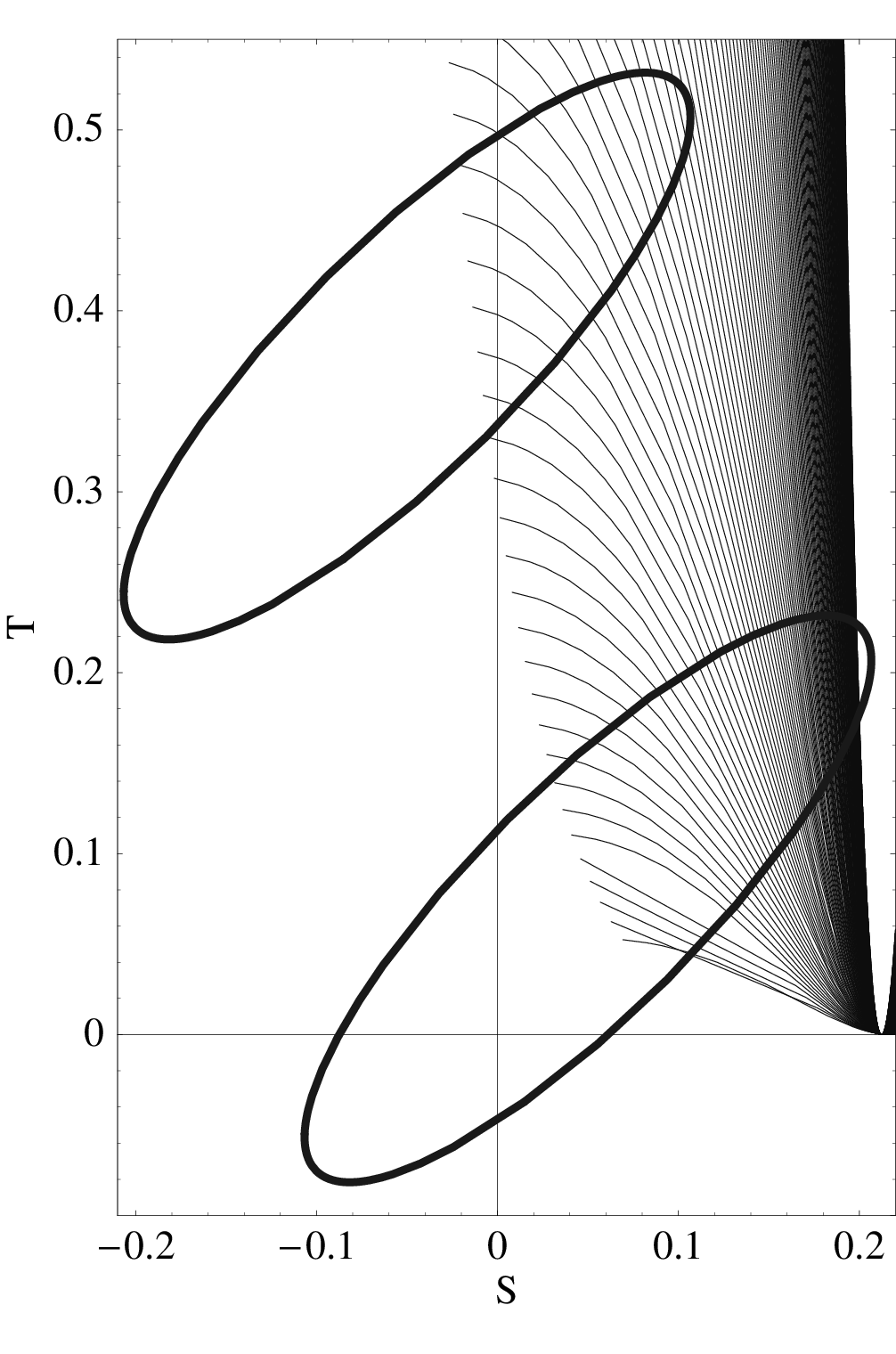}}
~~&~~~~
\resizebox{6.5cm}{!}{\includegraphics{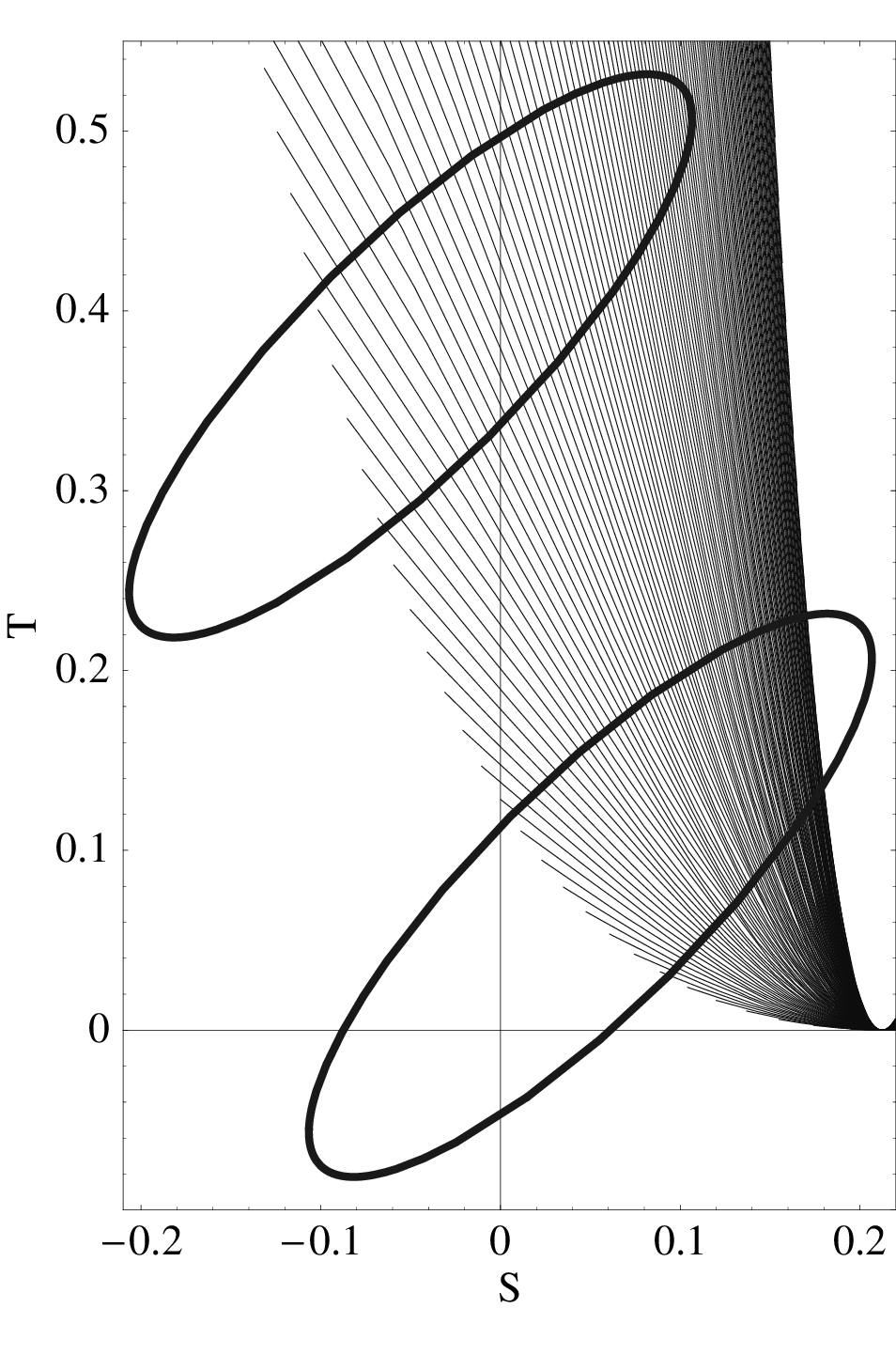}}
\end{tabular}
\caption{The ellipses represent the 68\% confidence region for the $S$ and $T$ parameters. The upper ellipse is for  a reference Higgs mass of the order of a TeV, the lower curve is for a light Higgs with mass around 114 GeV. The contribution from the MWT theory per se and of the leptons as function of the new lepton masses is expressed by the dark grey region. The left panel has been obtained using a SM type hypercharge assignment while the right hand graph is for $y=1$. }
\label{ST}
\end{figure}

Although we have already studied the effects of the lepton family on the
electroweak parameters in~\cite{Dietrich:2005jn}, we summarize here the main results
in Figure \ref{ST}.  The ellipses represent the 68\%
confidence region for the $S$ and $T$ parameters. The upper ellipse
is for  a reference Higgs mass of the order of 1 TeV while the lower
curve is for a light Higgs with mass around 114 GeV. The
contribution from the MWT theory per se and of the leptons as
function of the new lepton masses is expressed by the dark grey
region. The left panel has been obtained using a SM type hypercharge assignment while the right hand graph is for $y=1$. In both pictures the regions of overlap between the theory and the precision contours are achieved when the upper component of the weak isospin doublet is lighter than the lower component. The opposite case leads to a total $S$ which is larger than the one predicted within the new strongly coupled dynamics per se.  This is due to the sign of the hypercharge for the new leptons. The mass range used in the plots, in the case of the SM hypercharge assignment is $100-1000$~GeV for the new electron and $50-800$~GeV for the new Dirac neutrino, while it is $100-800$ and $100-1000$~GeV respectively for the $y=1$ case. The plots have been obtained assuming a Dirac mass for the new neutral lepton (in the case of a SM hypercharge assignment). The analysis for the Majorana mass case has been performed in \cite{Kainulainen:2006wq} where one can again show that it is easy to be within the 68\% contours.

The contour plots we have drawn take into account the new values of the top
mass which has dropped dramatically since we last compared our
theory in \cite{Dietrich:2005jn} to the experimental data \cite{EWWG}. 

\section{Conclusions}

We have provided a comprehensive extension of the standard model which embodies (minimal) walking technicolor theories and their interplay with the standard model particles. Our extension of the standard model features all of the relevant low energy effective degrees of freedom linked to our underlying minimal walking theory. These include scalars, pseudoscalars as well as spin one fields. The bulk of the Lagrangian has been spelled out. The link with underlying strongly coupled gauge theories has been achieved via the time-honored Weinberg sum rules. The modification of the latter according to walking has been taken into account. We have also analyzed the case in which the underlying theory behaves like QCD rather than being near an infrared fixed point. This has allowed us to gain insight on the spectrum of the spin one fields which is an issue of phenomenological interest. In the appendix we have: i) provided the explicit construction of all of the SU(4) generators, ii) shown how to construct the effective Lagrangian in a way which is amenable to quantum corrections,  iii) shown the explicit form of the mass matrices for all of the particles, iv) provided a summary of all of the relevant electroweak parameters and their explicit dependence on the coefficients of our effective theory.

We have introduced the model in a format which is, hopefully, {\it user  friendly} for collider phenomenology. 

\acknowledgments
We have benefitted from discussions with A. Belyaev, S. Catterall, D.D. Dietrich, C. Kouvaris, F. Krauss, J. Schechter and K. Tuominen. The work of R.F.,  M.T.F. and F.S. is supported by the Marie Curie Excellence Grant under contract MEXT-CT-2004-013510.

\appendix

\section{Generators\label{appgen}}

It is convenient to use the following representation of SU(4)
\beq S^a = \begin{pmatrix} \bf A & \bf B \\ {\bf B}^\dag & -{\bf A}^T
\end{pmatrix} \ , \qquad X^i = \begin{pmatrix} \bf C & \bf D \\ {\bf
    D}^\dag & {\bf C}^T \end{pmatrix} \ , \eeq
where $A$ is hermitian, $C$ is hermitian and traceless, $B = -B^T$ and
$D = D^T$. The ${S}$ are also a representation of the $SO(4)$
generators, and thus leave the vacuum invariant $S^aE + ES^T = 0\ $.
Explicitly, the generators read
\beq S^a = \frac{1}{2\sqrt{2}}\begin{pmatrix} \tau^a & \bf 0 \\ \bf 0 &
  -\tau^{aT} \end{pmatrix} \ , \quad a = 1,\ldots,4 \ , \eeq
where $a = 1,2,3$ are the Pauli matrices and $\tau^4 =
\mathbbm{1}$. These are the generators of SU$_V$(2)$\times$ U$_V$(1).
\beq S^a = \frac{1}{2\sqrt{2}}\begin{pmatrix} \bf 0 & {\bf B}^a \\
{\bf B}^{a\dag} & \bf 0 \end{pmatrix} \ , \quad a = 5,6 \ , \eeq
with
\beq B^5 = \tau^2 \ , \quad B^6 = i\tau^2 \ . \eeq
The rest of the generators which do not leave the vacuum invariant are
\beq X^i = \frac{1}{2\sqrt{2}}\begin{pmatrix} \tau^i & \bf 0 \\
\bf 0 & \tau^{iT} \end{pmatrix} \ , \quad i = 1,2,3 \ , \eeq
and
\beq X^i = \frac{1}{2\sqrt{2}}\begin{pmatrix} \bf 0 & {\bf D}^i \\
{\bf D}^{i\dag} & \bf 0 \end{pmatrix} \ , \quad i = 4,\ldots,9 \ ,
\eeq
with
\beq\begin{array}{r@{\;}c@{\;}lr@{\;}c@{\;}lr@{\;}c@{\;}l}
D^4 &=& \mathbbm{1} \ , & \quad D^6 &=& \tau^3 \ , & \quad D^8 &=& \tau^1 \ , \\
D^5 &=& i\mathbbm{1} \ , & \quad D^7 &=& i\tau^3 \ , & \quad D^9 &=& i\tau^1
\ .
\end{array}\eeq

The generators are normalized as follows
\beq {\rm Tr}\left[S^aS^b\right] =\frac{1}{2}\delta^{ab}\ , \qquad \ , {\rm Tr}\left[X^iX^j\right] =
\frac{1}{2}\delta^{ij} \ , \qquad {\rm Tr}\left[X^iS^a\right] = 0 \ . \eeq

\section{Vector Mesons as Gauge Fields} \label{sec:hidden}
We show how to rewrite the vector meson Lagrangian in a gauge invariant way.  We assume the scalar sector to transform according to a given but otherwise arbitrary representation of the flavor symmetry group $G$. This is a straightforward generalization of the Hidden Local Gauge symmetry idea \cite{Bando:1984ej,Bando:1987br}, used in a similar context for the BESS models \cite{Casalbuoni:1995qt}. At the tree approximation this approach is identical to the one introduced first in \cite{Kaymakcalan:1984bz,Kaymakcalan:1983qq}.

\subsection{Introducing Vector Mesons}
Let us start with a generic flavor symmetry group $G$ under which a scalar field $M$ transforms globally in a given, but generic, irreducible representation $R$.  We also introduce an algebra valued  one-form $A=A^{\mu}dx_{\mu}$ taking values in a copy of the algebra of the group $G$, call it $G^{\prime}$, i.e.
\begin{eqnarray}
 A_{\mu}=A^{a}_{\mu} T^a \ , \qquad {\rm with } \qquad T^a \in {\cal{A}}(G^{\prime}) \ .
 \end{eqnarray}
At this point the full group structure is the semisimple group $G\times G^{\prime}$. $M$ does not transform under $G^{\prime}$. Given that $M$ and $A$ belong to two different groups we need another field to connect the two. We henceforth introduce a new scalar field $N$ transforming according to the fundamental of $G$ and to the antifundamental of $G^{\prime}$. We then upgrade $A$ to a gauge field over $G^{\prime}$.
\begin{table}[t]
\caption{Field content}
\begin{center}
\begin{tabular}{c||cc}
&\,\,\,$G$\,\,\,&\,\,\,$G^{\prime}$\,\,\, \\
\hline
&&\\
$M$\,\,\,&\,\,\,$R$\,\,\,&\,\,\,${\mathbf 1}$\,\,\, \\
&&\\
$N$\,\,\,&\,\,\,$\fund$\,\,\,&\,\,\,$\overline{\fund}$\,\,\, \\
&&\\
$A_{\mu}$\,\,\,&\,\,\,${\mathbf 1}$\,\,\,&\,\,\,${\rm Adj}$\,\,\, \\
\end{tabular}
\end{center}
\label{default}
\end{table}%
The covariant derivative for $N$ is:
\begin{eqnarray}
D_{\mu}N=\partial_{\mu} N + i\,\tilde{g} \, N\,A_{\mu} \ .
\end{eqnarray}
We now force $N$ to acquire the following vacuum expectation value
\begin{eqnarray}
\langle N^i_j \rangle = \delta^i_j \, v^{\prime} \ ,
\end{eqnarray}
which leaves the diagonal subgroup - denoted with $G_{V}$ - of $G \times G^{\prime}$ invariant. Clearly $G_V$ is a copy of $G$. Note that it is always possible to arrange a suitable potential term for $N$ leading to the previous pattern of symmetry breaking. $v/v^{\prime}$ is expected to be much less than one and the {\it unphysical} massive degrees of freedom associated to the fluctuations of $N$ will have to be integrated out. The would-be Goldstone bosons associated to $N$ will become the longitudinal components of the massive vector mesons.

To connect $A$ to $M$ we define the one-form transforming only under $G$ via $N$ which - in the deeply spontaneously broken phase of $N$ - reads:
\begin{eqnarray}
\frac{ {\rm Tr} [N N^{\dagger}]}{{\rm dim}(F)} \, P_{\mu} =\frac{D_{\mu}N N^{\dagger} - N D_{\mu}N^{\dagger}} {2\, i \tilde{g}} \ , \qquad P_{\mu} \rightarrow u P_{\mu} u^{\dagger} \ ,
\end{eqnarray}
with $u$ being an element of $G$ and dim($F$) the dimension of the fundamental representation of $G$. When evaluating $P_{\mu}$ on the vacuum expectation value for $N$ we recover $A_{\mu}$:
\begin{eqnarray}
\langle P_{\mu } \rangle =  A_{\mu} \ .
\end{eqnarray}
  At this point it is straightforward to write the Lagrangian containing $N$, $M$ and $A$ and their self-interactions.
 Being in the deeply broken phase of $G\times G^{\prime}$ down to $G_{V}$ we count $N$ as a dimension zero field. This is consistent with the normalization for $P_{\mu}$.

The kinetic term of the Lagrangian is:
\begin{eqnarray}
L_{kinetic} = -\frac{1}{2}{\rm Tr} \left[F_{\mu\nu}F^{\mu\nu}\right]  + \frac{1}{2}{\rm Tr} \left[DN DN^{\dagger}\right] +\frac{1}{2}{\rm Tr} \left[ \partial M \partial M^{\dagger} \right] \ .
\end{eqnarray}
The second kinetic term will provide a mass to the vector mesons.
Besides the potential terms for $M$ and $N$ there is another part of the Lagrangian which is of interest to us. This is the one mixing $P$ and $M$.  Up to dimension four and containing at most two powers of $P$ and $M$ this is:
\begin{eqnarray}
L_{P-M} & = & \tilde{g}^2\ r_1 \ {\rm Tr}\left[P_\mu P^\mu M M^\dagger\right]
+ \tilde{g}^2\ r_2 \ {\rm Tr}\left[P_\mu M {P^\mu}^T M^\dagger \right] \nonumber \\
& + & i \ \tilde{g}\ r_3 \ {\rm Tr}\left[P_\mu \left(M (D^\mu M)^\dagger - (D^\mu M) M^\dagger \right) \right]
+ \tilde{g}^2\ s \ {\rm Tr}\left[P_\mu P^\mu \right] {\rm Tr}\left[M M^\dagger \right] \ .
\end{eqnarray}
The dimensionless parameters $r_1$, $r_2$, $r_3$, $s$ parameterize the strength of the interactions between the composite scalars and vectors in units of $\tilde{g}$, and are therefore expected to be of order one. We have assumed $M$ to belong to the two index symmetric representation of a generic G= SU(N). It is straightforward to generalize the previous terms to the case of an arbitrary representation $R$ with respect to any group G.
Further higher derivative interactions including $N$ can be included systematically.

\subsection{Further Gauging of G}
In this case we add another gauge field $G_{\mu}$ taking values in the algebra of $G$. We then define the correct covariant derivatives for $M$ and $N$. {}For $N$, for example, we have:
\begin{eqnarray}
D_{\mu}N=\partial_{\mu} N -i\,g\,G_{\mu}\,N+ i\,\tilde{g} \, N\,A_{\mu} \ .
\label{covariant-N}
\end{eqnarray}
Evaluating the previous expression on the vacuum expectation value of $N$ we recover the field $C_{\mu}$ introduced in the text. To be more precise we need to use $P_{\mu}$ again but with the covariant derivative for $N$ replaced by the one in the equation above.
\begin{table}[hb]
\caption{Field content}
\begin{center}
\begin{tabular}{c||cc}
&\,\,\,$G$\,\,\,&\,\,\,$G^{\prime}$\,\,\, \\
\hline
&&\\
$M$\,\,\,&\,\,\,$R$\,\,\,&\,\,\,${\mathbf 1}$\,\,\, \\
&&\\
$N$\,\,\,&\,\,\,$\fund$\,\,\,&\,\,\,$\overline{\fund}$\,\,\, \\
&&\\
$A_{\mu}$\,\,\,&\,\,\,${\mathbf 1}$\,\,\,&\,\,\,${\rm Adj}$\,\,\, \\
&&\\
$G_{\mu}$\,\,\,&\,\,\,${\rm Adj}$\,\,\,&\,\,\,${\mathbf 1}$\,\,\, \\
\end{tabular}
\end{center}
\label{default2}
\end{table}%

\section{Effective Lagrangian and Mass Matrices} \label{app:lagr}
In this section we summarize and generalize the effective Lagrangians for the scalar and vector sectors, and include the explicit mass matrices for the mixings of the composite vectors with the fundamental gauge fields. 
\subsubsection{Scalar Sector}
The composite scalars are assembled in the matrix $M$ of Eq.~(\ref{M}). In terms of the mass eigenstates this reads
\small
\begin{eqnarray}
M=
\begin{pmatrix}
i \Pi_{UU} + \widetilde{\Pi}_{UU} & {\displaystyle \frac{i \Pi_{UD} + \widetilde{\Pi}_{UD}}{\sqrt{2}}} &
{\displaystyle \frac{\sigma+i\Theta + i \Pi^0 + A^0}{2}} & {\displaystyle \frac{i\Pi^+ + A^+}{\sqrt{2}}} \\
\smallskip \\
{\displaystyle \frac{i \Pi_{UD} + \widetilde{\Pi}_{UD}}{\sqrt{2}}} & i \Pi_{DD} + \widetilde{\Pi}_{DD} &
{\displaystyle \frac{i\Pi^- + A^-}{\sqrt{2}}} & {\displaystyle \frac{\sigma+i\Theta - i \Pi^0 - A^0}{\sqrt{2}}} \\
\smallskip \\
{\displaystyle \frac{\sigma+i\Theta + i \Pi^0 + A^0}{2}} & {\displaystyle \frac{i\Pi^- + A^-}{\sqrt{2}}} &
i \Pi_{\overline{UU}} + \widetilde{\Pi}_{\overline{UU}} & {\displaystyle \frac{i \Pi_{\overline{UD}} + \widetilde{\Pi}_{\overline{UD}}}{\sqrt{2}}} \\
\smallskip \\
{\displaystyle \frac{i\Pi^+ + A^+}{\sqrt{2}}} & {\displaystyle \frac{\sigma+i\Theta - i \Pi^0 - A^0}{2}} &
{\displaystyle \frac{i \Pi_{\overline{UD}} + \widetilde{\Pi}_{\overline{UD}}}{\sqrt{2}}} & i \Pi_{\overline{DD}} + \widetilde{\Pi}_{\overline{DD}}
\end{pmatrix} \ , \nonumber \\
\end{eqnarray}
\normalsize
where $\sigma=v+H$. The Lagrangian for the Higgs sector, including the spontaneously broken potential, and the ETC mass term for the uneaten Goldstone bosons, is
\begin{eqnarray}
{\cal L}_{\rm Higgs} &=& \frac{1}{2}{\rm Tr}\left[D_{\mu}M D^{\mu}M^{\dagger}\right] + \frac{m^2}{2}{\rm Tr}[MM^{\dagger}] \nonumber \\
& - & \frac{\lambda}{4} {\rm Tr}\left[MM^{\dagger} \right]^2 - \lambda^\prime {\rm Tr}\left[M M^{\dagger} M M^{\dagger}\right]
+  2\lambda^{\prime\prime} \left[{\rm Det}(M) + {\rm Det}(M^\dagger)\right] \nonumber \\
& + & \frac{m_{\rm ETC}^2}{4}\ {\rm Tr}\left[M B M^\dagger B + M M^\dagger \right] \ ,
\end{eqnarray}
where the covariant derivative is given by Eq.~(\ref{covariantderivative}).

\subsubsection{Vector Sector}
In terms of the charge eigenstates the matrix $A^\mu$ is
\begin{eqnarray}
A^\mu =
\begin{pmatrix}
{\displaystyle \frac{a^{0\mu}+v^{0\mu}+v^{4\mu}}{2\sqrt{2}}} & 
{\displaystyle \frac{a^{+\mu}+v^{+\mu}}{2}} &  
{\displaystyle \frac{x_{UU}^\mu}{\sqrt{2}}} & 
{\displaystyle \frac{x_{UD}^\mu+s_{UD}^\mu}{2}} \\
\smallskip \\
{\displaystyle \frac{a^{-\mu}+v^{-\mu}}{2}} & 
{\displaystyle \frac{-a^{0\mu}-v^{0\mu}+v^{4\mu}}{2\sqrt{2}}} & 
{\displaystyle \frac{x_{UD}^\mu-s_{UD}^\mu}{2}} & 
{\displaystyle \frac{x_{DD}^\mu}{\sqrt{2}}} \\
\smallskip\\
{\displaystyle \frac{x_{\overline{UU}}^\mu}{\sqrt{2}}} & 
{\displaystyle \frac{x_{\overline{UD}}^\mu-s_{\overline{UD}}^\mu}{2}} & 
{\displaystyle \frac{a^{0\mu}-v^{0\mu}-v^{4\mu}}{2\sqrt{2}}} & 
{\displaystyle \frac{a^{-\mu}-v^{-\mu}}{2}} \\
\smallskip \\
{\displaystyle \frac{x_{\overline{UD}}^\mu+s_{\overline{UD}}^\mu}{2}} & 
{\displaystyle \frac{x_{\overline{DD}}^\mu}{\sqrt{2}}} & 
{\displaystyle \frac{a^{+\mu}-v^{+\mu}}{2}} & 
{\displaystyle \frac{-a^{0\mu}+v^{0\mu}-v^{4\mu}}{2\sqrt{2}}}
\end{pmatrix} \ .
\end{eqnarray}
The most general Lagrangian for the gauge and vector fields can be conveniently written using the $N$ and $P_\mu$ fields of appendix~\ref{sec:hidden}. Demanding $CP$ invariance, and including terms up to dimension four, we have 
\begin{eqnarray}
{\cal L}_{\rm vector} =  & - & \frac{1}{2}{\rm Tr} \Big[\widetilde{W}_{\mu\nu}\widetilde{W}^{\mu\nu}\Big]
-\frac{1}{4}{\rm Tr}\Big[B_{\mu\nu}B^{\mu\nu}\Big]
-\frac{1}{2}{\rm Tr} \Big[F_{\mu\nu}F^{\mu\nu}\Big] \nonumber \\
&+& \frac{1}{2}{\rm Tr} \left[D_\mu N \left(D^\mu N\right)^{\dagger}\right] 
+ \frac{1}{2}{\rm Tr} \left[D_\mu M \left(D^\mu M\right)^\dagger \right] \nonumber \\
& + &  \tilde{g}^2 \ a_1 \ {\rm Tr}\Big[P_\mu P^\mu \Big]^2 + \tilde{g}^2  \ a_2 \ {\rm Tr}\Big[P_\mu P^\mu P_\nu P^\nu\Big]
+ \tilde{g}^2 \ a_3 \ {\rm Tr}\Big[P_\mu P_\nu P^\mu P^\nu \Big]  \nonumber \\
& - & i\ \tilde{g} \ b \ {\rm Tr}\Big[[P_\mu,P_\nu] N F^{\mu\nu} N^\dagger \Big] \nonumber \\
& + & \tilde{g}^2\ r_1 \ {\rm Tr}\Big[P_\mu P^\mu M M^\dagger\Big]
+ \tilde{g}^2\ r_2 \ {\rm Tr}\Big[P_\mu M {P^\mu}^T M^\dagger \Big] \nonumber \\
& + & i \ \tilde{g}\ r_3 \ {\rm Tr}\Big[P_\mu \left(M (D^\mu M)^\dagger - (D^\mu M) M^\dagger \right) \Big]
+ \tilde{g}^2\ s \ {\rm Tr}\Big[P_\mu P^\mu \Big] {\rm Tr}\Big[M M^\dagger \Big] \ , \nonumber \\
\end{eqnarray}
where the field strength tensor $F^{\mu\nu}$ is given by Eq.~(\ref{strength}), and the covariant derivatives of $M$ and $N$ are respectively given by Eq.~(\ref{covariantderivative}) and Eq.~(\ref{covariant-N}). Notice that we have excluded the  terms $i\, {\rm Tr}\left[[P_\mu,P_\nu]G^{\mu\nu}\right]$ and ${\rm Tr}\left[N F_{\mu\nu} N^\dagger G^{\mu\nu}\right]$ with order one couplings.  There terms in the limit of no weak interactions are reserved solely to technicolor interactions.  Here $G^{\mu
\nu}$ contains $\widetilde{W}^{\mu\nu}$ and $B^{\mu\nu}$. 

The covariant derivative terms give rise to mass terms for the charged and neutral vector bosons:
\begin{eqnarray}
{\cal L}_{\rm mass} =
\begin{pmatrix} \widetilde{W}^-_\mu & v^-_\mu & a^-_\mu  \end{pmatrix} {\bf {\cal M}_{\rm C}^2}
\begin{pmatrix} \widetilde{W}^{+\mu} \\ v^{+\mu} \\ a^{+\mu}  \end{pmatrix} +
\frac{1}{2}\begin{pmatrix} B_\mu & \widetilde{W}^3_\mu & v^0_\mu & a^0_\mu & v^4_\mu \end{pmatrix} {\bf {\cal M}_{\rm N}^2}
\begin{pmatrix} B^\mu \\ \widetilde{W}^{3\mu} \\ v^{0\mu} \\ a^{0\mu} \\ v^{4\mu} \end{pmatrix} \ ,
\end{eqnarray}
where
\begin{eqnarray}
{\bf {\cal M}_{\rm C}^2} =
\begin{pmatrix}
{\displaystyle \frac{g^2 M_V^2 (1+\omega)}{\tilde{g}^2}} & 
-{\displaystyle \frac{g M_V^2}{\sqrt{2}\tilde{g}}} & 
-{\displaystyle \frac{g M_A^2 (1-\chi)}{\sqrt{2}\tilde{g}}} \\
\smallskip \\
-{\displaystyle \frac{g M_V^2}{\sqrt{2}\tilde{g}}} & 
M_V^2 &
0 \\
\smallskip \\
-{\displaystyle \frac{g M_A^2 (1-\chi)}{\sqrt{2}\tilde{g}}} & 
0 & 
M_A^2
\end{pmatrix} \ ,
\end{eqnarray}
\begin{eqnarray}
{\bf {\cal M}_{\rm N}^2} =
\begin{pmatrix}
{\displaystyle \frac{g^{\prime 2}M_V^2 (1+2y^2+\omega)}{\tilde{g}^2}} & 
-{\displaystyle \frac{g g^\prime M_V^2 \omega}{\tilde{g}^2}} & 
-{\displaystyle \frac{g^\prime M_V^2}{\sqrt{2}\tilde{g}}} &
{\displaystyle \frac{g^\prime M_A^2 (1-\chi)}{\sqrt{2}\tilde{g}}} & 
-{\displaystyle \frac{g^\prime M_V^2 (2y)}{\sqrt{2}\tilde{g}}} \\
\smallskip \\
-{\displaystyle \frac{g g^\prime M_V^2\omega}{\tilde{g}^2}}  &
{\displaystyle \frac{g^2 M_V^2 (1+\omega)}{\widetilde{g}^2}} &
-{\displaystyle \frac{g M_V^2}{\sqrt{2}\tilde{g}}} &
-{\displaystyle \frac{g M_A^2 (1-\chi)}{\sqrt{2}\tilde{g}}} &
0 \\
\smallskip \\
-{\displaystyle \frac{g^\prime M_V^2}{\sqrt{2}\tilde{g}}} &
-{\displaystyle \frac{g M_V^2}{\sqrt{2}\tilde{g}}} &
M_V^2 &
0 &
0 \\
\smallskip \\
{\displaystyle \frac{g^\prime M_A^2 (1-\chi)}{\sqrt{2}\tilde{g}}} &
-{\displaystyle \frac{g M_A^2 (1-\chi)}{\sqrt{2}\tilde{g}}} &
0 &
M_A^2 &
0 \\
\smallskip \\
-{\displaystyle \frac{g^\prime M_V^2 (2y)}{\sqrt{2}\tilde{g}}} &
0 &
0 &
0 &
M_V^2
\end{pmatrix}  \ . \nonumber \\
\end{eqnarray}
Here $M_V$ and $M_A$ are the masses of the vector and axial vector bosons in absence of electroweak interactions, and are related by
\begin{eqnarray}
M_A^2 = M_V^2 + \frac{1}{2}v^2\tilde{g}^2 r_2 \ .
\end{eqnarray}
The parameters $\omega$ and $\chi$ are defined by
\begin{eqnarray}
\omega & \equiv & \frac{v^2\tilde{g}^2}{4 M_V^2} (1+r_2-2r_3) \nonumber \\
\chi & \equiv & \frac{v^2\tilde{g}^2}{2 M_A^2} r_3 \ ,
\end{eqnarray}
where $\chi$ has been used already in Eq.~(\ref{S}).

The vector baryons do not mix with the fundamental gauge fields and thus their masses do not receive tree-level electroweak corrections. Therefore, $x_{UU}$, $x_{UD}$, and $x_{DD}$ are axial mass eigenstates, and $s_{UD}$ is a vector mass eigenstate:
\begin{eqnarray}
& M_{x_{UU}} & = M_{x_{UD}} = M_{x_{DD}} = M_A  \ , \nonumber \\
& M_{s_{UD}} & = M_V \ . 
\end{eqnarray}

\section{Universal Electroweak Corrections}
\label{electrocorrections}
Any extension of the Standard Model cannot be at odds with the precision electroweak data.
Universal corrections to the Standard Model, i.e. corrections to the self-energies of
the vector bosons, may be encoded in a total of
7 parameters following reference \cite{Barbieri:2004qk}.  We show the relation with the ones presented in the main text \cite{Peskin:1990zt} and use our newly built effective theory to explicitly evaluate the corrections within the MWT. Let $Q^2\equiv -q^2$ be the Euclidean transferred momentum, and denote derivatives with respect to $-Q^2$ with a prime. Then we have the following definitions \cite{Barbieri:2004qk}: 
\begin{eqnarray}
\hat{S} &\equiv & g^2 \ \Pi_{W^3B}^\prime (0) \ , \\
\hat{T} &\equiv & \frac{g^2}{M_W^2}\left[ \Pi_{W^3W^3}(0) -
\Pi_{W^+W^-}(0) \right] \ , \\
W &\equiv & \frac{g^2M_W^2}{2} \left[\Pi^{\prime\prime}_{W^3W^3}(0)\right] \ , \\
Y &\equiv & \frac{g'^2M_W^2}{2} \left[\Pi^{\prime\prime}_{BB}(0)\right] \ , \\
\hat{U} &\equiv & -g^2 \left[\Pi^\prime_{W^3W^3}(0)-
\Pi^\prime_{W^+W^-}(0)\right]\ , \\
V &\equiv & \frac{g^2 \, M^2_W}{2}\left[\Pi^{\prime\prime}_{W^3W^3}(0)-
\Pi^{\prime\prime}_{W^+W^-}(0)\right] \ , \\
X &\equiv & \frac{g g'\,M_W^2}{2} \ \Pi_{W^3B}^{\prime\prime}(0) \ .
\end{eqnarray}
Here $\Pi_V(Q^2)$ with $V=\{W^3B,\, W^3W^3,\, W^+W^-,\, BB\}$ represents the
self-energy of the vector bosons. Here the
electroweak couplings are the ones associated to the physical electroweak gauge bosons:
\begin{eqnarray}
\frac{1}{g^2} \equiv  \Pi^\prime_{W^+W^-}(0)
 \ , \qquad \frac{1}{g'^2}
\equiv  \Pi^\prime_{BB}(0) \ ,
\end{eqnarray}
while $G_F$ is
\begin{eqnarray}
\frac{1}{\sqrt{2}G_F}=-4\Pi_{W^+W^-}(0) \ ,
\end{eqnarray}
as in \cite{Chivukula:2004af}. $\hat{S}$ and $\hat{T}$ lend their name
from the well known Peskin-Takeuchi parameters $S$ and $T$ which are related to the new ones via
\cite{Barbieri:2004qk,Chivukula:2004af}:
\begin{eqnarray}
\frac{\alpha S}{4s_W^2} =  \hat{S} - Y - W  \ , \qquad 
\alpha T = \hat{T}- \frac{s_W^2}{1-s_W^2}Y \ .
\end{eqnarray}
Here $\alpha$ is the electromagnetic structure constant and $s_W^2$
is the weak mixing angle. Therefore in the case where $W=Y=0$ we
have the simple relation
\begin{eqnarray}
\hat{S} &=& \frac{\alpha S}{4s_W^2} \ , \qquad 
\hat{T}= \alpha T \ .
\end{eqnarray}
In our model these parameters read:
\begin{eqnarray}
\hat{S} &=& \frac{(2 - \chi )\chi g^2}{2\tilde{g}^2+(2-2\chi+\chi^2)g^2}  
      \ , \\ 
\hat{T} &=& 0 \ , \\
W &=& M_W^2 \frac{g^2(M_A^2+(\chi-1)^2M_V^2)}{(2\tilde{g}^2+(2+(\chi-2
)\chi)g^2)M_A^2M_V^2} \ ,  \\
Y &=& M_W^2 \frac{g'^2(5M_A^2+(\chi -1)^2M_V^2)}{(2\tilde{g}^2+(6+(\chi
-2)\chi)g'^2)M_A^2M_V^2}
 \ , \\
\hat{U} &=& 0 \ , \\
V &=& 0 \ , \\
X &=& g\,g' \, \frac{M_W^2}{M^2_A M^2_V}\frac{ M_A^2 - (\chi -1)^2M_V^2 }{\sqrt{(2\tilde{g}^2+(2-2\chi+\chi^2)g^2)(2\tilde{g}^2+(6-2\chi+\chi^2)g'^2)}} 
 \ .
\end{eqnarray}
In these expressions the coupling constants $g$, $g'$ and $\tilde{g}$ are the ones in the  Lagrangian associated to the yet to be diagonalized spin one states. $W$, $Y$ and $X$ are sensitive to the ratio $M^2_{W}/M^2$ with $M^2$ the lightest of the massive spin one fields. We have checked that even taking an axial vector mass as small as 500~GeV while keeping large $\tilde{g}$ for fixed $S$ of order $0.1$ one is able to satisfy the experimental constraints on all of the parameters.

\end{document}